\documentclass[aps,jcp,superscriptaddress,amsmath,amssymb]{revtex4-1}

\usepackage{comment}
\usepackage{tikz}
\usetikzlibrary{trees}
\usetikzlibrary{decorations.pathmorphing}
\usetikzlibrary{decorations.markings}
\usepackage{tikz-cd}
\usetikzlibrary{arrows,arrows.meta}
\usetikzlibrary{decorations.pathreplacing,snakes}
\tikzset{
    photon/.style={decorate, decoration={snake,segment length=1.5mm}, draw=black},
    coulomb/.style={dotted},
    electron/.style={draw=black, postaction={decorate},
        decoration={markings,mark=at position .55 with {\arrow[draw=black]{>}}}}, 
    gluon/.style={decorate, draw=magenta,
        decoration={coil,amplitude=4pt, segment length=5pt}},
    boundelectron/.style={thick, double},
    transverse/.style={dashed},
    marrow/.style={decoration={markings,mark=at position 0.5 with {\arrow{#1}}}, postaction=decorate}    
}
\usepackage{graphicx}
\usepackage{dcolumn}
\usepackage{bm}
\usepackage{rotating}
\usepackage{siunitx}
\newcolumntype{.}{D{.}{.}{8}}

\usepackage[utf8]{inputenc}
\usepackage{physics}
\usepackage{amsmath, amssymb}
\usepackage{tabularx,booktabs,array}
\usepackage{setspace}
\usepackage{vmargin}
\usepackage{xcolor}
\usepackage{graphicx,psfrag,subfigure}

\usepackage{float}
\usepackage[all]{xy}

\usepackage{bm}
\usepackage{mathtools}
\usepackage{physics}
\usepackage[english]{babel}

\newcolumntype{d}[1]{D{.}{.}{#1}}

\newcommand{\bos}[1]{\boldsymbol{#1}}
\newcommand{\mr}[1]{\mathrm{#1}}

\def\Eh{E_\mathrm{h}}

\def\iim{\mr{i}}
\def\eem{\mr{e}}

\def\four{^{[4]}} 
\def\sixteen{^{[16]}} 

\def\bp{\boldsymbol{p}}
\def\br{\boldsymbol{r}}

\def\bs{\boldsymbol{s}}

\def\bsigma{\boldsymbol{\sigma}}

\def\hmcA{{\mathcal{A}}}
\def\hmcS{{\mathcal{S}}}
\def\hH{{h}}
\def\hone{{1}}
\def\hP{{P}}
\def\pone{{(1)}}
\def\ptwo{{(2)}}

\def\cutting{\text{cutting}}

\def\punching{\text{punching}}

\def\honehtwo{$h_1h_2$}

\usepackage[unicode]{hyperref}
\usepackage{soul}
\hypersetup{
   unicode=true,          
   plainpages=false,
   colorlinks=true,       
   linkcolor=black,          
   linkcolor=blue,          
   citecolor=blue,        
   urlcolor=blue           
}

\usepackage{url}

\begin{document}

\title{%
One-Particle Operator Representation over Two-Particle Basis Sets for Relativistic QED Computations
}

\author{Péter Hollósy}
\author{Péter Jeszenszki}
\author{Edit M\'atyus} 
\email{edit.matyus@ttk.elte.hu}
\affiliation{ELTE, Eötvös Loránd University, Institute of Chemistry, 
Pázmány Péter sétány 1/A, Budapest, H-1117, Hungary}

\date{\today}

\begin{abstract}
\noindent %
This work is concerned with two-spin-1/2-fermion relativistic quantum mechanics, and it is about the construction of one-particle projectors using an inherently two(many)-particle, `explicitly correlated' basis representation, necessary for good numerical convergence of the interaction energy.
It is demonstrated that a faithful representation of the one-particle operators, which appear in intermediate but essential computational steps, can be constructed over a many-particle basis set by accounting for the full Hilbert space beyond the physically relevant anti-symmetric subspace.
Applications of this development can be foreseen for the computation of quantum-electrodynamics corrections for a correlated relativistic reference state and high-precision relativistic computations of medium-to-high-$Z$ helium-like systems, for which other two-particle projection techniques are unreliable.
\end{abstract}

\maketitle


\section{Introduction}
Recent developments of precision spectroscopy experimental techniques \cite{BeHoHuChSaEiUbJuMe19,SeJaCaMeScMe20,ClJaScAgScMe21,GuBaPRHoCa21,ShBaReHoCa23,ClScAgScMe23,test21,AlGiCoKoSc20,sci2020} have triggered interest in the computation of the relativistic energy by direct solution of the Dirac equation instead of computing increasingly high orders and increasingly complicated perturbation theory corrections to the non-relativistic energy \cite{Ye01,KoYe01,Pa06,KoTs07,KoHiKa13,KoHiKa14,PaYePa19,PaYePa20,PaYeVlPa21,YePaPa22}. 
High-precision computation of the Dirac relativistic energy automatically carries high (all) orders of relativistic corrections. In particular, basis set methods \cite{NoKa23} as well as finite-element techniques \cite{KuSc22} were recently developed to compute high-precision eigenstates of the one-electron Dirac equation for H$_2^+$-like two-center systems to catch up with the increasing experimental accuracy. 
Precision spectroscopy and precision physics experiments are becoming available also for poly-electronic systems \cite{SeJaCaMeScMe20,ClJaScAgScMe21,ClScAgScMe23}, the ongoing (two-electron) triplet helium puzzle is an illustrative example \cite{ClJaScAgScMe21,PaYeVlPa21,ClScAgScMe23}.

Regarding the theoretical framework for two-electron and two-spin-1/2 fermion systems, we have recently elaborated \cite{MaFeJeMa23,MaMa24,NoMaMa24} the Bethe--Salpeter equation \cite{SaBe51} and its equal-time variant pioneered by Salpeter \cite{Sa52} and Sucher \cite{sucherPhD1958} (see also Araki \cite{araki57}). The no-pair approximation to the equal-time Bethe--Salpeter(--Sucher) equation results in the no-pair Dirac--Coulomb(--Breit), in short, DC(B) Hamiltonian. 
A high-precision solution of the no-pair DC(B) equation can be achieved by using explicitly correlated basis functions \cite{ByPeKa08,PeByKa06,PeByKa07,PeByKa12,Ka17,JeFeMa21,JeFeMa22,FeJeMa22,FeJeMa22b,JeMa23,FeMa23}, similarly to the non-relativistic Schrödinger case \cite{SuVa98,rmp13,MaRe12,JeIrFeMa22,RoJeMaPo23,FeMa22h3}.
To construct the no-pair Hamiltonian, not only the Hamiltonian matrix elements but also the projector states must be constructed during the computations. It turns out that the computation of the (positive-energy) projectors is non-trivial in an explicitly correlated framework.
Although the projector states are based on one-electron properties and do not contain any information about the electron-electron interaction, if we used an orbital-based projector, the accuracy of the orbital representation would limit the precision of the solution of the interacting eigenvalue problem. 
For this reason, Li, Shao, and Liu proposed \cite{LiShLi12} to use a large, auxiliary one-electron basis set for the projector representation. 
Bylicki, Pestka, and Karwowski developed an inherently two-electron projector over an explicitly correlated basis set by relying on specific properties of the complex-coordinate rotated (CCR) two-electron Dirac Hamiltonian \cite{ByPeKa08}.
The choice of the projector has been the subject of recent research \cite{AlKnJeDySa16} considering also old arguments from Mittleman \cite{Mi81}.

In our recent explicitly relativistic computations, we used the CCR projector and a simple energy-cutting scheme. Although both approaches worked well and allowed us to converge the no-pair Dirac--Coulomb(--Breit) energy to more than 8 significant digits \cite{JeFeMa21,JeFeMa22,FeJeMa22b,JeMa23}, none of them was without problems, especially beyond the lowest $Z$ values. Furthermore, the computation of the quantum electrodynamics corrections to the correlated relativistic energy, which is ongoing work in our research group \cite{MaFeJeMa23,MaMa24,NoMaMa24}, requires a systematic computational approach to the construction of the different non-interacting, two-electron subspaces ($++$, $+-$, $-+$, or $--$, Fig.~\ref{fig:pmsubspaces}), which motivated the development of an inherently one-particle scheme while working with non-separable two-particle basis functions.

\section{An explicitly correlated no-pair Dirac--Coulomb(--Breit) computational approach in a nutshell \label{sec:nopairsum}}
The Dirac Hamiltonian for a spin-1/2 particle with mass $m_a$ is 
\begin{align}
  \tilde{h}_a^{[4]}
  =
  c(\boldsymbol{\alpha}^{[4]}\cdot\boldsymbol{p}_a)
  +
  \beta^{[4]} m_a c^2 + U_aI^{[4]} \; ,
\quad %
  h_a^{[4]} 
  = 
  \tilde{h}_a^{[4]}-I^{[4]} m_a c^2 \; ,
  \label{eq:singlepartDirac}
\end{align}
where $a = 1,2$ is the index of the particle and we have also defined the operator shifted with $m_ac^2$ to match the non-relativistic energy scale. The $\boldsymbol{\alpha}^{[4]}$ and $\beta^{[4]}$ Dirac matrices have their usual definition,
\begin{align}
    \boldsymbol{\alpha}^{[4]}
    =  
    \left(%
    \begin{array}{cc}
        \boldsymbol{0}^{[2]} & \boldsymbol{\sigma}^{[2]} \\
       \boldsymbol{\sigma}^{[2]}   &  \boldsymbol{0}^{[2]} 
    \end{array}\right) \; ,
    \hspace{2cm}  
    \beta^{[4]} = \left( \begin{array}{cc}
        I^{[2]} & 0^{[2]} \\
        0^{[2]}   &  -I^{[2]} 
    \end{array}\right) \ 
    \label{eq:diracmx}
\end{align}
with the $\bos{\sigma}^{[2]}=(\sigma_x^{[2]},\sigma_y^{[2]},\sigma_z^{[2]})$ Pauli matrices and the $I^{[n]}$ $n$-dimensional unit matrix. $c$ is the speed of light, which is the inverse of the fine-structure constant in atomic units, $c=1/\alpha$.
$\boldsymbol{p}_a$ labels the momentum operator, and $U_a$ carries the interaction potential energy due to an external electromagnetic field, \emph{e.g.,} the Coulomb interaction energy of the electrons with the clamped atomic nuclei,
\begin{align}
    U_a = -\sum_A \frac{Z_A}{\left| \bos{r}_a - \bos{R}_A \right|} \ ,
\end{align}
where $\bos{r}_a$ is the position of the particle (electron) $a$, $Z_A$ and $\bos{R}_A$ are the charge number and position of the nucleus $A$.

\begin{figure}
\begin{center}
\includegraphics[scale=0.6]{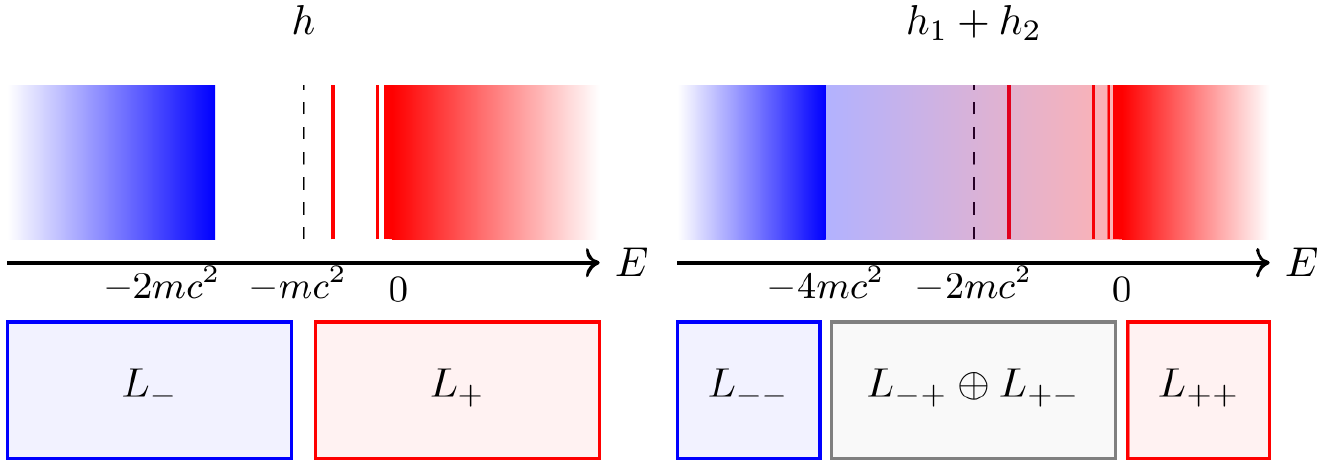}
\end{center}
\caption{%
  Visual representation of the one-particle $+$, $-$, and the two-particle
  $++$, $+-$, $-+$, and $--$ subspaces of the one- and two-particle Dirac Hamiltonians.
  \label{fig:pmsubspaces}
}
\end{figure}

By solving the Dirac equation,
\begin{align}
    h_a^{[4]} \left| \varphi_{ia}^{(4)} \right \rangle = e_{ia} \left| \varphi_{ia}^{(4)} \right \rangle \; ,
\end{align}
the $e_{ia}$ single-particle energies can be separated into two branches according to the lowest-energy bound-state solution and the large ($\sim 2m_a c^2$) gap in the energy spectrum. The states in the higher(lower)-energy branch constitute the so-called positive(negative)-energy space,
\begin{align}
  L_{a+}^{[4]} 
  = 
  \sum_i \left| \varphi_{ia}^{+{(4)}} \right \rangle {\left \langle \varphi_{ia}^{+{(4)}} \right|} \ ,   
  \hspace{2cm} 
  L_{a-}^{[4]} 
  = 
  \sum_i \left| \varphi_{ia}^{-{(4)}} \right \rangle {\left \langle \varphi_{ia}^{-{(4)}} \right|} \ ,
\end{align}
where $ \left| \varphi_{ia}^+ \right \rangle$ ($ \left| \varphi_{ia}^- \right \rangle$) has a corresponding $e_{ia}^+$ ($e_{ia}^-$) single-particle energy from the upper (lower) branch, 
and the $i$ index runs over both the continuous and discrete part of the spectra.

For two spin-1/2 particles (electrons), the no-pair approximation to the equal-time Bethe--Salpeter equation \cite{sucherPhD1958,MaFeJeMa23,MaMa24,NoMaMa24} provides a linear eigenvalue equation with the Hamiltonian, 
\begin{align}
 H^{[16]}=L_{++}^{[16]}\left(h^{[4]}_1\boxtimes I^{[4]}+I^{[4]}\boxtimes h^{[4]}_2 + V I^{[16]}+b J^{[4]} \otimes B^{[4]} \right)L_{++}^{[16]} \; ,
 \label{eq:ham16}
\end{align}
where $\otimes$ is the usual direct product, and $\boxtimes$ is a block-wise direct product which allows us to retain some of the quantities from the one-particle theory \cite{ShLiLi17,TrSi72}, Eqs.~\eqref{eq:singlepartDirac}--\eqref{eq:diracmx}.
In particular, $L_{++}^{[16]}=L_{1+}^{[4]} \boxtimes L_{2+}^{[4]}$ projects onto the positive-energy space of both  particles. The $V$ Coulomb interaction energy including the $r_{12}$ distance of the electrons,
\begin{align}
  V =  \frac{1}{r_{12}} \ 
 \end{align}
appears along the diagonal of $H^{[16]}$, whereas the Breit interaction is in the anti-diagonal blocks \cite{FeJeMa22,FeJeMa22b}, according to 
\begin{align}
J^{[4]}=\left(\begin{array}{cccc}
      0 & 0 & 0 & 1 \\
     0 & 0 & 1 & 0 \\
     0 & 1 & 0 & 0 \\
     1 & 0 & 0 & 0
 \end{array} \right) \ , \hspace{1cm}B^{[4]}=\
  -\frac{1}{2}
    \sum\limits_{i=1}^3\sum\limits_{j=1}^3 
      \left[%
      \frac{\delta_{ij}}{r_{12}}+\frac{1}{2} \left\lbrace \grad_{1i}\grad_{2j}r_{12} \right\rbrace 
      \right]
      \sigma^{[4]}_{1i} \sigma^{[4]}_{2j}  
\end{align}
with $\sigma_1^{[4]}=\sigma^{[2]}\boxtimes I^{[2]}$ and $\sigma_2^{[4]}=I^{[2]}\boxtimes \sigma^{[2]}$.

In Eq.~\eqref{eq:ham16}, $b$ is introduced for a compact definition of the Dirac--Coulomb (DC) ($b=0$) and Dirac--Coulomb--Breit (DCB) ($b=1$) Hamiltonians. 

The block-wise form of $H^{[16]}$ is
{\footnotesize %
\begin{align}
  H\sixteen
  =
  L_{++}^{[16]}
  \left(%
  \begin{array}{@{}cccc@{}}
    \left(V  + U \right)I\four & 
    c \bsigma_2\four \bp_2 &
    c \bsigma_1\four \bp_1 &
    bB\four \\
    c \bsigma_2\four \bp_2 &
    \left(V + U-2m_2 c^2 \right)I\four &
    bB\four &
    c\bsigma_1\four \bp_1 \\
    c\bsigma_1\four \bp_1 & 
    bB\four &
    \left(V + U-2 m_1 c^2 \right) I\four &
    c\bsigma_2\four \bp_2 \\
    bB\four & 
    c\bsigma_1\four \bp_1 &
    c\bsigma_2\four \bp_2 &
    \left( V + U-2m_{12}c^2 \right) I\four \\
  \end{array}
  \right) 
  L_{++}^{[16]}
  \ ,
  \label{eq:ham16block}
\end{align}
}
where $m_{12}=m_1+m_2$, $U=U_1+U_2$. We also note that an energy shift is also introduced in the single-particle part of the diagonal elements for straightforward comparison with the non-relativistic energy scale (Fig.~\ref{fig:pmsubspaces}). 

The no-pair energy, $E$, and wave function, $\Psi^{(16)}$, are determined by the eigenvalue equation,
\begin{align}
  H^{[16]} \left| \Psi_i^{(16)} \right \rangle 
  = 
  E_i  \left| \Psi_i^{(16)} \right \rangle \ .
\end{align}

For the numerical solution of this eigenvalue equation, $\Psi^{(16)}$ is expanded over a sixteen-dimensional spinor basis set \cite{JeFeMa21,JeFeMa22,FeJeMa22,FeJeMa22b,JeMa23},
\begin{align}
   \left|  \Psi_i^{(16)} \right \rangle  
   &=     
     \sum_{{n}=1}^{N_\text{b}} \sum_{q=1}^{16} 
       c_{{n}q,i} 
       P_G^{\zeta {[16]}} \mathcal{A}^{[16]} X^{[16]} | \Phi_{{n}q}^{(16)} \rangle  \ , 
   \label{eq:ansatz}
\end{align}
where $c_{{n}q,i}$ is the linear expansion coefficient, $\mathcal{A}^{[16]}$ is the anti-symmetrization operator for the two electrons \cite{JeFeMa22}, and $P_G^{\zeta [16]}$ is a projector corresponding to the $\zeta$ irreducible representation (irrep) of the $G$ point group \cite{DyFaBook07,JeMa23}.  

Furthermore, we implement the basis representation of $H^{[16]}$ using the $X^{[16]}$ kinetic balance matrix \cite{Ku84,SuLiKu11,JeFeMa22,FeJeMa22,FeJeMa22b},
\begin{align}
  X\sixteen
  =
  \left(%
  \begin{array}{@{}cccc@{}}
    1\four & 
    0\four &
    0\four &
    0\four \\
    0\four &
    \frac{\bsigma_2\four\bp_2}{2m_2c} &
    0\four &
    0\four \\
    0\four & 
    0\four &
    \frac{\bsigma_1\four\bp_1}{2m_1c} &
    0\four \\
    0\four & 
    0\four & 
    0\four &
    \frac{(\bsigma_1\four \bp_1)(\bsigma_2\four \bp_2)}{4m_1m_2c^2} \\
  \end{array}
  \right) \; . \label{eq:kinbal}
\end{align}
$X^{[16]}$ automatically ensures necessary spatial symmetry relations, and it is understood as part of the basis set definition. 

In our explicitly correlated computations, we have implemented $X^{[16]}$ as a transformation of the Hamiltonian, $X^{\dagger[16]} H\sixteen X\sixteen$ \cite{JeFeMa22,FeJeMa22}, as well as the identity operator, which gives rise to the $X^{\dagger[16]} X^{[16]} $ metric. The detailed expressions can be found in Refs.~\citenum{JeFeMa22} and \citenum{FeJeMa22}.

A $| \Phi_{{n}q}^{(16)} \rangle$ sixteen-component basis function can be defined as 
\begin{align}
    | \Phi_{{n}q}^{(16)} \rangle  &= I_q^{(16)}  \, | \Theta_{n} \rangle \  \label{eq:basis}
\end{align}
with $(I_q^{(16)})_p= \delta_{pq}$  (in the elementary spinor representation). 
$\Theta_{n}(\br)$ is a floating explicitly correlated Gaussian (fECG) function \cite{SzJe10,SuVa98,MaRe12,Ma19review}, 
\begin{align}
  \Theta_{n}(\br) 
  &= 
  \exp\left[%
    -\left( \br-\bs_{n} \right)^\text{T} 
    \underline{\bos{A}}_{n} 
    \left( \br - \bs_{n} \right)
  \right] \ ,
  \label{eq:ecg}
\end{align}
where the $\underline{\bos{A}}_{n}=\bos{A}_{n}\otimes I^{[3]}$  positive-definite exponent matrix with $\bos{A}_{n}\in\mathbb{R}^{2\times 2}$ and the $\bs_{n}\in\mathbb{R}^{3\times 2}$ `shift' vector are optimized variationally by minimization of the non-relativistic energy. In this work, numerical results are reported within the singlet basis sector (in the $LS$ coupling scheme for atoms), which dominates the no-pair energy of low-$Z$ systems. Triplet basis functions can be added according to Ref.~\citenum{JeMa23}, and their leading-order contribution to the no-pair energy is at $\alpha^4\Eh$ order in excellent numerical agreement with non-relativistic QED.

\section{Positive-energy projectors}\label{sec:Proj}
In order to construct the matrix representation of $H^{[16]}$, Eq.~\eqref{eq:ham16}, we must be able to deal with the $L^{[16]}_{++}$ projection `operator'. A formal definition of the two-electron projector can be written as
\begin{align}
  L^{[16]}_{++}=L^{[4]}_{1+} \boxtimes L^{[4]}_{2+} 
  \quad\text{with}\quad
  L^{[4]}_{a+} 
  = 
  \frac{%
    \tilde{h}^{[4]}_a+|\tilde{h}^{[4]}_a|
  }{%
    2|\tilde{h}^{[4]}_a|
  }
  \; ,
  \label{eq:projabs}
\end{align}
where the absolute value of the Hamiltonian is understood as $|\tilde{h}^{[4]}_a| = \sum_i |\tilde{e}_{ia}||\varphi_{ia}^{(4)}\rangle\langle\varphi_{ia}^{(4)}|$ (note $\tilde{h}_a$ from Eq.~\eqref{eq:singlepartDirac}).

In orbital-based approaches, the construction of the one-particle projectors is straightforward \cite{DyFaBook07,ReWoBook15,Ku12}.
So, it may first sound like a practical idea to apply an orbital-based approach only for the construction of the positive-energy projector (`non-interacting space corresponding to positive energies') and use it in the full computation (also including the electron-electron interaction) with an explicitly correlated basis set \cite{JeFeMa22,Liu12,LiShLi17}. 
In this case, the most demanding part of the computation is the evaluation of the overlap matrix of the explicitly correlated basis set and a large, auxiliary orbital-based (determinant) basis. 
(An auxiliary basis set was first proposed by Liu \cite{Liu12} to allow for using a larger basis set for the projector.) 
Unfortunately, a closer look at this problem, as well as our numerical experience, shows that
even with a large auxiliary basis set, the precision of the no-pair energy is limited by the size of the auxiliary orbital basis, and any benefit from using an explicitly correlated basis set is lost.

In essence, a faithful projector (matrix) representation \emph{within} the subspace spanned by our explicitly correlated basis set is extremely important. So, we must be able to construct the projectors beyond an orbital representation, \emph{i.e.,} over inherently `two-particle' basis states.

Computational strategies for the construction of the (positive-energy) projector over an explicitly correlated basis space have been developed in the past (including our work), but none of them is without problems. In what follows, we provide a short overview of each technique and explain the known deficiencies.

\subsection{Energy cutting projection technique}
The energy-cutting technique starts with the computation of the eigenstates of the two-electron, non-interacting, bare (unprojected) Hamiltonian over the explicitly correlated basis set, 
\begin{align}
  H^{[16]}_0
  =
  h^{[4]}_1\boxtimes I^{[4]}+I^{[4]}\boxtimes h^{[4]}_2  \; .
  \label{eq:nonintH}
\end{align}
Then, the non-interacting energy levels are arranged in increasing energetic order, and only those states are retained in further computations for which the energy is larger than a certain energy threshold ($E_\text{th}$). This threshold energy is a (tight) lower bound to the non-interacting ground-state energy. 
This computational technique was conceived \cite{JeFeMa21,JeFeMa22,FeJeMa22} as a quick, preparatory check before the more involved complex-coordinate rotation (CCR) \emph{(vide infra)} projection \cite{ByPeKa08} is carried out. The no-pair Hamiltonian (matrix) constructed with this projector is hermitian, but the cutting projector contains contaminant BR ($+-$ and $-+$) states with energy larger than the $E_\text{th}$ threshold, and for this reason, it is not a rigorous projection technique.

For the compact, variationally optimized ECG basis sets used throughout our work \cite{JeFeMa21,JeFeMa22,FeJeMa22,FeMa23,JeMa24CC,MaFeJeMa23}, the simple cutting technique was found to work surprisingly well. For the extensively studied low-$Z$ systems, the no-pair (cutting) energies are in excellent agreement with the more rigorous no-pair (CCR) energies, 
and the $\alpha$ fine-structure-constant dependence of the no-pair energy is in excellent numerical agreement with the relevant non-relativistic QED (nrQED) values, currently in use as golden standard theory reference for precision spectroscopy \cite{WaYa18,FeMa19EF,FeMa19HH,FeKoMa20,SaFeMa22,PuKoCzPa16}. 

Even more, we have conjectured that the simple `cutting projector' provided  numerical results superior to the CCR projector \emph{(vide infra)} for medium-$Z$ systems. At the same time, we were aware of the fundamental limitations of the cutting projection technique and have considered the CCR projector in principle better, but less useful in practical (ECG) computations.

\subsection{Complex-coordinate rotation and energy punching projection techniques \label{sec:ccrpunch}}
The complex coordinate rotation (CCR) technique is based on the different behaviour of the positive- and negative-energy branches of the one-electron  Hamiltonian upon the $\br \rightarrow \br \eem^{\iim\vartheta}$ complex scaling, \emph{i.e.,} complex-coordinate rotation, of the particle's coordinates for atoms \cite{ByPeKa08,JeFeMa22} and also molecules \cite{JeFeMa22}
\begin{align}
  \br_{a} \rightarrow \br_{a}\eem^{\iim\vartheta}: %
  \quad %
  h_{a}^{[4]}\rightarrow \bar{h}_{a}^{[4]}(\vartheta) 
  &= 
  c(\boldsymbol{\alpha}^{[4]}\cdot\boldsymbol{p}_{a})\eem^{-\iim \vartheta}
  +
  (\beta^{[4]}-I^{[4]})m_{a} c^2
  +
  \bar{U}_{a}(\vartheta)I^{[4]} \; ,
  \label{eq:hCCR}  
  \\
  \bar{U}_{a}(\vartheta) 
  &= 
  -\sum_A \frac{Z_A}{\left| \bos{r}_a\eem^{\iim\vartheta} - \bos{R}_A \right|} \; .
\end{align}
The eigenvalues corresponding to the $h \approx c(\boldsymbol{\alpha}^{[4]}\cdot\boldsymbol{p})\eem^{-\iim \vartheta}+(\beta^{[4]}-I^{[4]})m c^2$ approximation, \emph{i.e.,} large momentum limit with negligible contribution from the external potential, are \cite{Se88} 
\begin{align}
 \lim_{\left|e \right| \rightarrow \infty} \bar{e} (e, \vartheta) = \begin{cases} +\sqrt{\eem^{-2\iim\vartheta} e^2+m^2 c^4} -mc^2 \approx  +\eem^{-\iim\vartheta} e-\frac{1}{2}m^2c^4 
 \hspace{0.5cm} \mbox{for} \hspace{0.3cm} e \gg 0 \ , \\ 
 -\sqrt{\eem^{-2\iim\vartheta} e^2+m^2 c^4} -mc^2 \approx  -\eem^{-\iim\vartheta} e-\frac{3}{2}m^2c^4 
 \hspace{0.5cm} \mbox{for} \hspace{0.3cm} e \ll 0 \ ,
 \end{cases}
  \label{eq:ECCRlin}
\end{align}
which shows the qualitatively different `rotation' of the positive- and negative-energy branches (about different `centers', Fig.~\ref{fig:proj}) in the complex plane upon changing $\vartheta$.

This behaviour was first exploited by Bylicki, Pestka, and Karwowski \cite{ByPeKa08} to separate the positive-energy ($L_{++}$) eigenfunctions of the
the two-electron, non-interacting Hamiltonian, 
\begin{align}
  H^{[16]}_0(\vartheta)
  =
  h^{[4]}_1(\vartheta)\boxtimes I^{[4]}+I^{[4]}\boxtimes h^{[4]}_2(\vartheta)  \; ,
  \label{eq:nonintHCCR}
\end{align}
represented over an explicitly correlated basis set.
Due to the qualitatively different $\vartheta$ trajectories of the eigenvalues from the $+$ and $-$ branches, Eq.~\eqref{eq:ECCRlin}, the $++$, the Brown--Ravenhall (BR: $+-$, $-+$), and the $- -$ subspaces can be distinguished even in a non-separable (explicitly correlated) basis set. 

In principle, the identification of the $++$ states can be performed for any $\vartheta$ rotation angle, which is sufficiently large for a clear distinction of the different branches. 
Then, for the selected $\vartheta$ angle, the (CCR transformed) no-pair Hamiltonian matrix can be constructed and its eigenvalues are the no-pair energies. The no-pair energies of the bound states (including the physically relevant ground state) are real (within basis convergence), but the no-pair-CCR-Hamiltonian is non-hermitian, so the variational upper-bound property of the energy is lost and one has to work with the left- and right-handed no-pair-CCR eigenfunctions in further (\emph{e.g.,} perturbation theory) computations. 

\paragraph{A punching projector?}
In principle, it should be possible to eliminate the non-hermitian feature of the no-pair (CCR) Hamiltonian, by tracking the $++$ (BR and $--$) branches back to the $\vartheta\rightarrow 0$ limit. It is important to note that since we are interested in the computation of bound states, any small $\vartheta$ value may be appropriate for which the different branches can be clearly (numerically) separated.
`Back-rotating' the CCR branches  ($\vartheta\rightarrow 0$) to the real axis would allow us to identify the BR states that contaminate our energy cutting list (above the $E_\text{th}$ threshold energy and retained for positive-energy computations with the cutting projector). In principle, this technique, named `punching projection' \cite{JeFeMa22} (we punch a `hole' in the cutting energy list where the contaminating BR state is identified), would combine the rigour of the CCR projector and the hermiticity of the cutting projector. 

Unfortunately, for medium-$Z$ nuclear charge numbers, 
we noticed ambiguities in separating the different branches of the $H_0^{[16]}(\vartheta)$ Hamiltonian ($Z=18$ subfigure in Fig.~\ref{fig:proj}). 

These ambiguities are absent for the proposed \honehtwo\ projection approach, by construction.

\section{Consecutive projection to the one-particle spaces\label{sec:h1h2proj}}
To rigorously eliminate the positive-energy BR space while using the Hilbert space spanned by the explicitly correlated basis functions, we propose a two-step projection scheme for a two-spin-1/2 fermion system. 
In the first step, we identify the positive-energy subspace of the $h_1$ Hamiltonian (`1+') within the Hilbert space spanned by the basis functions, and then, \emph{within this 1+ subspace}, we select the positive-energy subspace of $h_2$, which defines the `1+2+=++' subspace.  Alternatively, we could start with $h_2$, and then, continue with $h_1$, resulting in the same ++ subspace \emph{(vide infra)}.

But how to construct the matrix representation of $h_1$ and $h_2$ over a non-separable basis space?! Aren't they the same matrices, half of the matrix representation of the non-interacting two-particle Hamiltonian, $(h_1+h_2)/2$? Well, to have a faithful representation of a \emph{one-particle quantity over a two-particle space,} we must use the \emph{entire Hilbert space}, beyond its (physically relevant) anti-symmetric subspace. So, we construct the $\bos{h}_1$ and $\bos{h}_2$ matrices over two-particle basis functions, and both the anti-symmetrized and the symmetrized two-particle functions must be included. (Alternatively, we could simply work with a non-symmetrized spinor-spatial basis set.)

We define the permutational antisymmetrization and symmetrization operators as
\begin{align}
  \hmcA = \frac{1}{2}(\hone-\hP_{12})
  \quad\text{and}\quad
  \hmcS = \frac{1}{2}(\hone+\hP_{12}) \; ,
\end{align}
which are idempotent, $\hmcA\hmcA=\hmcA$ and $\hmcS\hmcS=\hmcS$, and thus, 
\begin{align}
  \left[%
  \begin{array}{@{}c|c@{}}
    \hmcA \hH_1^{{[16]}} \hmcA & \hmcA \hH_1^{{[16]}} \hmcS \\
    \hline
    \hmcS \hH_1^{{[16]}} \hmcA & \hmcS \hH_1^{{[16]}} \hmcS \\
  \end{array}
  \right]
  &=
  \frac{1}{2}
  \left[%
  \begin{array}{@{}c|c@{}}
    \hmcA (\hH_1^{{[16]}}+\hH_2^{{[16]}}) \hmcA & 
    \hmcA (\hH_1^{{[16]}}-\hH_2^{{[16]}}) \hmcS \\
    \hline
    \hmcS (\hH_1^{{[16]}}-\hH_2^{{[16]}} \hmcA & 
    \hmcS (\hH_1^{{[16]}}+\hH_2^{{[16]}}) \hmcS \\
  \end{array}
  \right] \; ,
\end{align}
\begin{align}
  \left[%
  \begin{array}{@{}c|c@{}}
    \hmcA \hH_2^{{[16]}} \hmcA & 
    \hmcA \hH_2^{{[16]}} \hmcS \\
    \hline
    \hmcS \hH_2^{{[16]}} \hmcA & 
    \hmcS \hH_2^{{[16]}} \hmcS \\
  \end{array}
  \right]
  =
  \frac{1}{2}
  \left[%
  \begin{array}{@{}c|c@{}}
    \hmcA (\hH_1^{{[16]}}+\hH_2^{{[16]}}) \hmcA & 
    \hmcA (\hH_2^{{[16]}}-\hH_1^{{[16]}}) \hmcS \\
    \hline
    \hmcS (\hH_2^{{[16]}}-\hH_1^{{[16]}}) \hmcA & 
    \hmcS (\hH_1^{{[16]}}+\hH_2^{{[16]}}) \hmcS \\
  \end{array}
  \right] \; ,
\end{align}
and 
\begin{align}
  \left[%
  \begin{array}{@{}c|c@{}}
    \hmcA (\hH_1^{{[16]}}+\hH_2^{{[16]}}) \hmcA & 
    \hmcA (\hH_1^{{[16]}}+\hH_2^{{[16]}}) \hmcS \\
    \hline
    \hmcS (\hH_1^{{[16]}}+\hH_2^{{[16]}}) \hmcA & 
    \hmcS (\hH_1^{{[16]}}+\hH_2^{{[16]}}) \hmcS \\
  \end{array}
  \right]
  =
  \left[%
  \begin{array}{@{}c|c@{}}
    \hmcA (\hH_1^{{[16]}}+\hH_2^{{[16]}}) \hmcA & 
    0^{{[16]}} \\
    \hline
    0^{{[16]}} & 
    \hmcS (\hH_1^{{[16]}}+\hH_2^{{[16]}}) \hmcS \\
  \end{array}
  \right]  \; .
\end{align}
The result of this simple calculation is, of course, textbook material, \emph{e.g.,} \cite{EyWaKi44,Lo55}.

Then, if we use the short notation for spinor basis functions in the permutationally anti-symmetric and symmetric subspaces by $a_k$ and $s_j$ ($k=1,\ldots,N_\text{A}$ and $j=1,\ldots,N_\text{S}$), we can build and diagonalize the $\bos{h}_1$ and $\bos{h}_2$ Hamiltonians as
\begin{align}
  \bos{h}_1 
  =
  \left(%
    \begin{array}{@{}c|c@{}}
      \langle a_k | {h}_1^{{[16]}} | a_i \rangle 
      & 
      \langle a_k | {h}_1^{{[16]}} | s_j \rangle \\
      \hline %
      \langle s_l | {h}_1^{{[16]}} | a_i \rangle
      & 
      \langle s_l | {h}_1^{{[16]}} | s_j \rangle 
    \end{array}
  \right)
  =
  \sum_{n=1}^{N_\text{A}+N_\text{S}}
    \lambda_n^\pone | \chi^\pone_n \rangle \langle \chi^\pone_n |
  \; ,
  \label{eq:H1mxas}
\end{align}
and 
\begin{align}
  \bos{h}_2
  =
  \left(%
    \begin{array}{@{}c|c@{}}
      \langle a_k | {h}_2^{{[16]}} | a_i \rangle 
      & 
      \langle a_k | {h}_2^{{[16]}} | s_j \rangle \\
      \hline %
      \langle s_l | {h}_2^{{[16]}} | a_i \rangle 
      & 
      \langle s_l | {h}_2^{{[16]}} | s_j \rangle \\
    \end{array}
  \right)
  =
  \sum_{n=1}^{N_\text{A}+N_\text{S}}
    \lambda_n^\ptwo | \chi^\ptwo_n \rangle \langle \chi^\ptwo_n |  \; .
  \label{eq:H2mxas}
\end{align}
$\lambda_n^\pone,\chi^\pone_n$ and
$\lambda_n^\ptwo,\chi^\ptwo_n$ ($n=1,\ldots,N_\text{A}+N_\text{S}$)
are distinct sets of eigenpairs. For the special basis parameterization of the ECGs with $\bos{s}=0$, Eq.~\eqref{eq:ecg},
$\lambda_n^\pone=\lambda_n^\ptwo$ and the corresponding $\chi^\pone_n$ and $\chi^\ptwo_n$ eigenfunctions are related, they differ only in their relative phase over the anti-symmetric and symmetric subspaces.

\begin{figure}
    \centering
    \includegraphics[scale=.75]{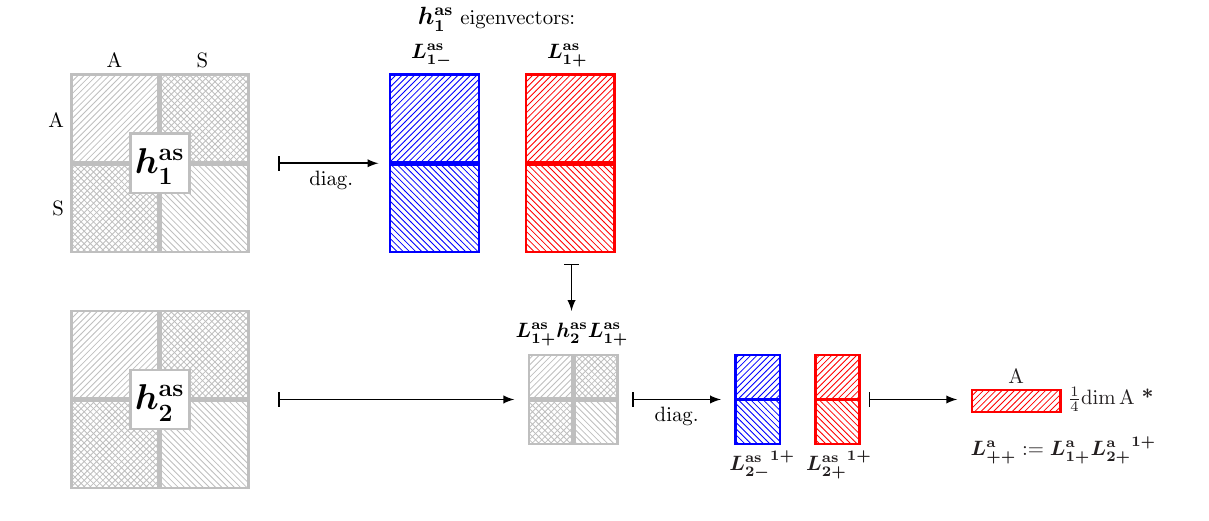}
    \caption{%
      Main steps of the construction of the $L_{++}$ positive-energy projector by consecutive diagonalization of the one-particle Hamiltonians over the explicitly correlated basis set.
      In the final step labelled with $^\ast$, the selected basis states are expressed as a linear combination of the anti-symmetrized elementary ECG spinor basis f, Eq.~\eqref{eq:ansatz}.
      %
    }
    \label{fig:Flowchart}
\end{figure}

To build the matrix representation of $h_1$ ($h_2$) over the explicitly correlated basis set, we consider the $h_1\sixteen$ (and $h_2\sixteen$) operator in which the identity over the second particle spinor space is explicitly written as
\begin{align}
  h_1\sixteen
  =
  h^{[4]}_1\boxtimes I^{[4]}
  =
  \left(%
  \begin{array}{@{}cccc@{}}
    U_1 1\four & 
    0\four &
    c \bsigma_1\four \bp_1 &
    0\four \\
    0\four &
    U_1 1\four &
    0\four &
    c\bsigma_1\four \bp_1 \\
    c\bsigma_1\four \bp_1 & 
    0\four &
    (U_1-2 m_1 c^2) 1\four &
    0\four \\
    0\four & 
    c\bsigma_1\four \bp_1 &
    0\four &
    (U_1-2m_1 c^2) 1\four \\
  \end{array}
  \right) \; ,
\end{align}
and similarly
\begin{align}
  h_2\sixteen
  =
  I^{[4]}\boxtimes h^{[4]}_2
  =
  \left(%
  \begin{array}{@{}cccc@{}}
    U_2 1\four & 
    c\bsigma_2\four \bp_2 & 
    0\four & 
    0\four \\
    c\bsigma_2\four \bp_2 & 
    (U_2 -2m_2 c^2) 1\four &
    0\four & 
    0\four \\
    0\four &
    0\four &
    U_2 1\four &
    c\bsigma_2\four \bp_2 \\
    0\four & 
    0\four &
    c\bsigma_2\four \bp_2 &
    (U_2 - 2 m_2 c^2) 1\four \\
  \end{array}
  \right) \; .
\end{align}
The matrix representations of the one-electron Hamiltonians are constructed, similarly to the two-electron Hamiltonian, using the two-particle kinetic balance, Eq.~\eqref{eq:kinbal}, implemented as a transformation or metric. Then, the $X$-transformed one-electron Hamiltonians are
\begin{align}
  &X^\dagger h_1\sixteen X
  \nonumber \\
  &=
  \left(%
  \begin{array}{@{}cccc@{}}
    U_1 1\four & 
    0\four &
    \frac{\bp_1^2}{2m_1} 1\four &
    0\four \\
    0\four &
    \frac{U_1 \bp_2^2}{4 m_2^2 c^2} 1\four &
    0\four &
    \frac{\bp_1^2\bp_2^2}{8 m_1 m_2^2 c^2} 1\four \\
    \frac{\bp_1^2}{2 m_1} 1\four & 
    0\four &
    \frac{(\bsigma_1\four \bp_1) U_1 1\four (\bsigma_1\four \bp_1)}{4m_1^2c^2} -\frac{\bp_1^2}{2m_1} 1\four &
    0\four \\
    0\four & 
    \frac{\bp_1^2 \bp_2^2}{8 m_1 m_2^2 c^2} 1\four & 
    0\four &
    \frac{(\bsigma_1\four \bp_1) U_1 1\four (\bsigma_1\four \bp_1)\bp_2^2}{16 m_1^2 m_2^2 c^4} - \frac{m_1 \bp_1^2 \bp_2^2}{8 m_1^2 m_2^2 c^2} 1\four
    \\
  \end{array}
  \right) \;   
\end{align}
and
\begin{align}
  &X^\dagger h_2\sixteen X
  \nonumber \\
  &=
  \left(%
  \begin{array}{@{}cccc@{}}
    U_2 1\four & 
    \frac{\bp_2^2}{2 m_2} 1\four & 
    0\four & 
    0\four \\
    \frac{\bp_2^2}{2 m_2} 1\four & 
    \frac{(\bsigma_2\four \bp_2)U_2 1\four (\bsigma_2\four \bp_2)}{4 m_2^2 c^2} - \frac{\bp_2^2}{2 m_2} & 
    0\four & 
    0\four \\
    0\four & 
    0\four & 
    \frac{\bp_1^2 U_2 1\four }{4 m_1^2 c^2} &
    \frac{\bp_1^2 \bp_2^2}{8 m_1^2 m_2 c^2} 1\four \\
    0\four & 
    0\four &
    \frac{\bp_1^2 \bp_2^2}{8 m_1^2 m_2 c^2} 1\four & 
    \frac{\bp_1^2 (\bsigma_2\four \bp_2) U_2 1\four (\bsigma_2\four \bp_2)}{16 m_1^2 m_2^2 c^4} - \frac{m_2 \bp_1^2 \bp_2^2}{8 m_1^2 m_2^2 c^2} 1\four
    \\
  \end{array}
  \right) \; .
\end{align}

Finally, we can check that the $h_1\sixteen+h_2\sixteen$ sum gives the non-interacting two-particle Hamiltonian $H_{12}\sixteen$ ($H\sixteen$ with $V=0$ and $B\four=0$), and the same identity also applies to the transformed Hamiltonians,
\begin{align}
  &X^\dagger h_1\sixteen X + X^\dagger h_2\sixteen X = X^\dagger H_{0}\sixteen X \; .
\end{align}
Ref.~\citenum{JeFeMa22} reports the two-electron expression, $X^\dagger H_{0}\sixteen X$, in detail.

This consecutive projection technique, which we call henceforth the `\honehtwo\ projection' for short, has been implemented in the 
in-house developed QUANTEN computer program according to the algorithmic steps shown in Fig.~\ref{fig:Flowchart}.
The procedure is formally and numerically invariant to the exchange of $h_1$ and $h_2$. In particular, identical no-pair energies are obtained if we first diagonalize $\bos{h}_1$, and then, $\bos{h}_2$ or \emph{vice versa.} 
We note that a similar construction of the $--$ subspace (retaining the negative-energy part in both one-electron steps) is also invariant to the actual ordering of the $h_1$ and $h_2$ diagonalization steps. At the same time, if we adopted the same procedure for the construction of the $+-$ space, the result would depend on the $h_1$ and $h_2$ order, likewise the construction of the $-+$ space (alone), but if we unify the $+-$ and $-+$, then the resulting BR ($+-$ and $-+$) subspace is invariant to the order of the $h_1$ and $h_2$ diagonalization steps. So, we plan to further develop the present procedure to avoid a prescribed ordering and to construct the intersection of the relevant $h_{1\sigma_1}$ and $h_{2\sigma_2}$ subspaces ($\sigma_1,\sigma_2=+$ or $-$) over the two-particle basis space. The procedure to compute the intersection of two vector spaces is often referred to as the `Zassenhaus algorithm'.

We also note that beyond two electrons, one has to consider the corresponding (larger) permutation group (of the identical particles) and include all irreducible representations (irreps), or simply work with asymmetric basis states, and then, project to the Pauli-allowed irrep (if needed) in the final step. 

Furthermore, the procedure (Fig.~\ref{fig:Flowchart}) can be straightforwardly adapted to two-spin-1/2 pre-Born--Oppenheimer, no-pair Dirac--Coulomb(--Breit) computations \cite{FeMa23}, as a fundamentally rigorous replacement to the simple cutting projector.


\section{Numerical results \label{sec:NumRes}}
\subsection{Characteristics of the projection techniques in action}

\begin{figure}
    \centering
    \includegraphics[scale=0.5]{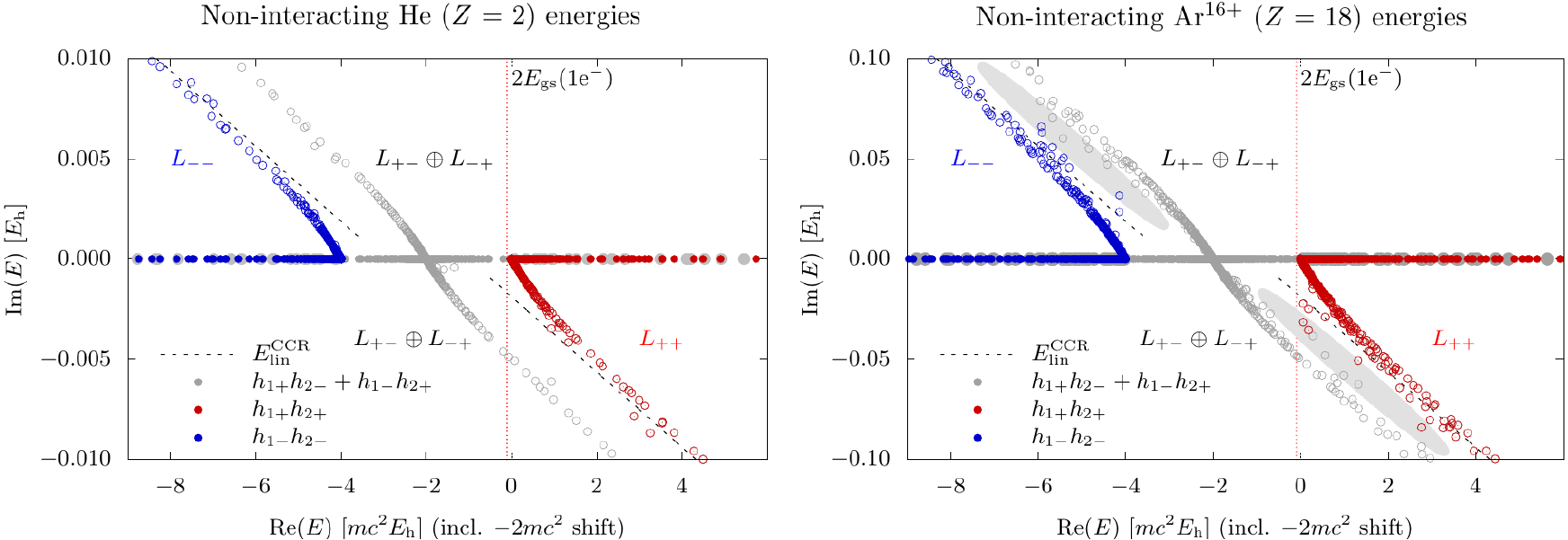}
\caption{%
    Non-interacting energies computed with 
    the CCR ({\color{blue}$\circ$}, {\color{gray}$\circ$}, {\color{red}$\circ$}) 
    and 
    the \honehtwo\ ({\color{blue}$\bullet$}, {\color{gray}$\bullet$}, {\color{red}$\bullet$}) 
    projection techniques. 
    The CCR energies (of non-bound states) are rotated in the complex plane 
    corresponding to the $\vartheta=10^{-7}$ CCR angle. 
    Twice the one-electron Dirac energy of the physical ground-state, $2E_\text{gs}(1\text{e}^-)$ is also shown, and a lower bound to this value is used as the threshold energy in the cutting projector.
%
    \label{fig:proj}
}
\end{figure}

Figure~\ref{fig:proj} highlights features of the CCR, the cutting, and the \honehtwo\ projection techniques for the examples of He ($Z=2$) and Ar$^{16+}$ ($Z=18$).

The CCR projector exploits the different `rotation' of the complex-scaled energies of the different branches in the complex plane with respect to the $\vartheta$ CCR angle, Eqs.~\eqref{eq:hCCR}--\eqref{eq:ECCRlin}. The linear energy dependence of the complex energy in the large momentum limit, Eq.~\eqref{eq:ECCRlin}, is also shown for the $++$ and  $--$ branches. 
An automated assignment of the non-interacting two-particle states to the $++$ ($--$) branch is performed based on the distance of their energy from this $++$ ($--$) limiting linear function.
It can be observed in the figure that the $Z=2$ states can be sorted into the different branches without problems, but ambiguities arise already for $Z=18$ (shown as an example of the problematic behaviour). By experimenting with the possible assignments of the states in the grey-shaded area, substantial variations in the no-pair energy can be observed (especially for even larger $Z$ values).
In the infinite basis limit, these ambiguities are expected to disappear, and we also note that Bylicki, Pestka, and Karwowski had a clearer separation of the branches in their Hyleraas CI procedure \cite{ByPeKa08}. With our compact ECG basis set, the no-pair DC(B) energies appear to converge to a well-defined value, and their $\alpha$
fine-structure constant dependence is consistent with nrQED values to high precision \cite{JeFeMa22,FeJeMa22b,JeMa23}, if the positive-energy projector can be assigned unambiguously.

The ambiguities experienced with the CCR projector (and ECGs, Fig.~\ref{fig:proj}) are absent for the newly proposed \honehtwo\ projector. The \honehtwo\ non-interacting, two-electron energies are plotted in Fig.~\ref{fig:proj} along the $x$ axis. The non-interacting states are automatically assigned to the ++, BR=$(+-,-+)$ and $--$ branches by construction (labelled in colour Fig.~\ref{fig:Flowchart}), and they are obtained in independent computations.

The colour coding of the energies along the $x$ (real) axis, available from the \honehtwo\ scheme, makes the fundamental deficiency of the cutting projector apparent. The cutting projector includes all non-interacting states for which the energy is larger than the predefined $E_\text{th}$ threshold. 
In the figure, we observe grey dots among the many red dots along the $x$ axis, which correspond to the positive-energy BR states contaminating the ++ space of the cutting projector. 

It is interesting to add that the back-rotated ($\vartheta\rightarrow 0$) CCR non-interacting energies (proposed for a punching(CCR) projector, if it can be unambiguously defined) are not identical with the \honehtwo\ energies, but they typically agreed to 5-6 digits in our computations. We can understand this small deviation originating from the different manipulations over a correlated basis set, and the deviations are expected to disappear in the infinite basis limit.

\begin{table}
  \caption{%
    No-pair Dirac--Coulomb energies, in $\Eh$, computed with the newly developed \honehtwo\ projector for the example 
    of atomic and molecular ground states (with clamped nuclei). 
    All converged digits and an additional 1-2 digits are shown.   
    Deviation of the energies from cutting projector results (using the same basis set),$^\ast$ $\delta E =E_\cutting-E_{\text{\honehtwo}}$, in $\Eh$, is also shown. 
    All deviations are smaller than the estimated basis convergence error of the energy.
    \label{tab:refdata}
  }
  \centering
  \begin{tabular}{@{}l ccc @{}}
    \hline\hline \\[-0.34cm]
    && $E_{\text{\honehtwo}}$ 
    & $\delta E$
    \\
    \hline \\[-0.34cm]
    He 
    && --2.903 856 631 6 & $5\cdot10^{-12}$ 
    \\
    Li$^+$
    && --7.280 698 894 5 & $9\cdot10^{-12}$ 
    \\
    Be$^{2+}$
    && --13.658 257 602 3 & --$6\cdot10^{-11}$ 
    \\
    Ar$^{16+}$
    && --314.246 104 2 & --$3\cdot10^{-8}$ 
    \\
    \hline \\
    H$_2$ 
    && --1.174 489 753 7 & --$6\cdot10^{-12}$
    \\
    H$_3^+$ 
    && --1.343 850 526 1 & $2\cdot10^{-12}$
    \\
    HeH$^{+}$ 
    && --2.978 834 635 4 & $3\cdot10^{-14}$ 
    \\
    \hline\hline
  \end{tabular}
  ~\\[-0.45cm]
  \begin{flushleft}
    $^\ast$: 
    The ECG basis set, taken from Refs.~\citenum{JeFeMa22,FeJeMa22b}, includes 700 (He), 400 (Li$^+$), 300 (Be$^{2+}$), 800 (Ar$^{16+}$), 800 (H$_2$), 400 (H$_3^+$), 1200 (HeH$^+$) functions. 
    The threshold energy in the cutting projector was the (analytically calculable) non-interacting energy for the atomic systems. For the molecular systems (no analytic value is known), and $E_\text{th}$ was within a 1~$\Eh$ lower bound to the numerically computed non-interacting energy, and its variation within this lower-bound window did not affect the no-pair energies shown in the table (Fig.~\ref{fig:CuttingThreshold}).
  \end{flushleft}
\end{table}

\subsection{Numerical application of the \honehtwo\ projector: computation of the no-pair energy}
As a first application of the \honehtwo\ projector, we computed the no-pair Dirac--Coulomb energy for a range of atomic and molecular systems. In Table~\ref{tab:refdata}, the \honehtwo\ no-pair DC energies are compared with the no-pair energies of former cutting projector results \cite{JeFeMa22,FeJeMa22b}. The deviation of the (fundamentally correct) \honehtwo\ and the (practical) cutting projector is very small, smaller than the basis set convergence error of the energy. 

We also note that we obtain higher (so, in a variational sense, worse) energies if we use the punching(CCR) projector constructed by following back the assigned CCR branches to the real axis ($\vartheta\rightarrow 0$). Alternatively, we can delete states from the cutting projector space based on the \honehtwo\ positive-energy list (by simple energy comparison), named punching(\honehtwo) projector. The no-pair energies obtained either with the punching(CCR) or the punching(\honehtwo) projectors are typically 2-3 orders of magnitude less accurate (in a variational sense) than the \honehtwo\ no-pair energies (in excellent agreement with the cutting projector results, Table~\ref{tab:refdata}). We can attribute this difference to a slower convergence rate of the no-pair (punching) energy for finite basis sizes.

\begin{figure}
    \centering
    \includegraphics[scale=1.05]{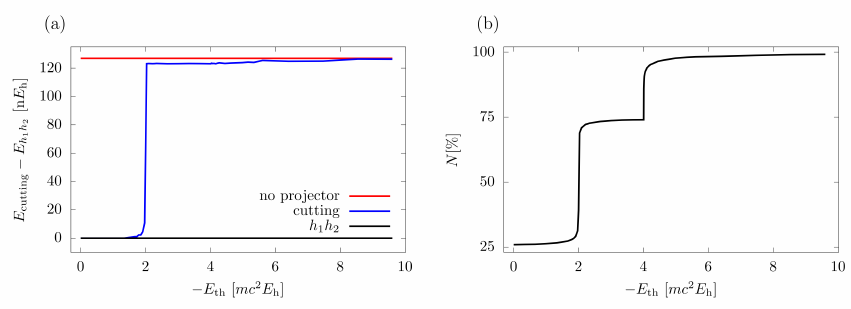}
    \caption{%
      Projected Dirac--Coulomb energy of the helium-atom ground state computed with the cutting projector as a function of the $E_\mathrm{th}$ threshold energy used to define the projector (a).
      The fraction of the retained basis states in the projector with respect to the total ($++$, $+-,-+$, $--$) spinor space is shown in subfigure (b).
      The number of ECG functions is $N = 700$ (Table~\ref{tab:refdata}) and $E_{\text{\honehtwo}}$ = --2.903 856 632~$E_\text{h}$.
      \label{fig:CuttingThreshold}      
      }
\end{figure}

\subsection{Why is the energy-cutting projector so good and when does it fail?}

For small to medium $Z$ values, the simple energy-cutting projector was found to perform extremely well in practical computations, although it suffers from (positive-energy) BR contamination, which would certainly manifest itself in the infinite basis limit. 

To better understand this numerical behaviour, we varied the $E_\text{th}$ threshold energy for the example of the helium atom ground state. Figure~\ref{fig:CuttingThreshold} shows the deviation of the no-pair Dirac--Coulomb energy as a function of the $E_\text{th}$ threshold energy, which is used in the definition of the cutting projector. There is a surprisingly long `plateau' at the beginning of the cutting energy curve (Fig~\ref{fig:CuttingThreshold}a), where its deviation from the \honehtwo\ no-pair energy is negligible (smaller than the basis set convergence error). Along this plateau, there is only a small increase in the number of contaminating states (Fig~\ref{fig:CuttingThreshold}b). 

Then, at around $-2mc^2$ (the `mid-point' of the BR energy range, also note the $-mc^2$ shift for both particles in Eq.~\eqref{eq:ham16block}), we observe a sudden jump, almost all the positive energy and BR states become part of the cutting projector (ca. 75~\% in Fig.~\ref{fig:CuttingThreshold}b), and the cutting energy jumps to the range very close to the `eigenvalue' of the bare (unprojected) interacting Hamiltonian. Then, by further lowering the $E_\text{th}$ value, the entire space becomes part of the `projector' (no projection happens) and we reach the eigenvalue of the unprojected Hamiltonian (at least a numerical approximation to its real part \cite{PeByKa06}). We note that the unprojected interacting Hamiltonian is considered to be problematic \cite{PeByKa06,PeByKa07,PeByKa12}, and it is only the positive-energy projected (no-pair) DC(B) Hamiltonian which has been derived from relativistic QED \cite{sucherPhD1958,Su80,Su83,Su84,MaFeJeMa23,MaMa24}.

So, Figure~\ref{fig:CuttingThreshold} highlights the robustness of the no-pair energy computed with the cutting projector with respect to small variations of the threshold energy near its physically motivated value, at least for the example of the ground state of the helium atom and an extensively optimized, compact ECG basis set.

It is also interesting to note in Fig.~\ref{fig:CuttingThreshold} that the projected relativistic energy \emph{increases} by including larger portions of the two-particle space in the `extended projector'. To better understand this behaviour, it is necessary to observe that if not the no-pair projector but the artificially extended projector (Fig.~\ref{fig:CuttingThreshold}) is used, then the physical state is obtained as (a highly) excited state in the energy list, and the energy of most BR states is by $-2mc^2E_\mathrm{h}$ less than the energy of the no-pair ground state. 
Then, by second-order perturbation theory, we can estimate the energy contribution to the no-pair ground state due to the low-energy BR states as 
\begin{align}
    E^{[2]} = \sum_k^{E_0>E_k>E_\mathrm{th}} \frac{\left| \left \langle \Psi_0 \left | V^{[16]} \right|  \Psi_k \right \rangle \right|^2 }{E_0-E_k} \ ,
\end{align}
where both the nominator and the denominator are positive, hence the negative-energy BR states increase the energy (higher-order perturbative corrections are negligible due to the large energy difference).

All in all, Fig.~\ref{fig:CuttingThreshold} highlighted the remarkable stability of the no-pair energy with respect to the choice of the threshold energy used to define the cutting projector.
Nevertheless, the cutting projector has a fundamental deficiency. If we increased the basis size approaching the infinite basis limit, Brown--Ravenhall continuum states would accumulate above the $E_\text{th}$ threshold energy and the lowest-energy `no-pair' state would be obtained an excited state (in the continuum starting at $E_\text{th}$).
We have not yet detected such an example during the extensive study of low-$Z$ systems (with a compact ECG basis) and $\alpha$ values near the CODATA18 $\alpha_0=1/137.035999084$ recommended value. So, we designed a stress test for the cutting projector to numerically observe this failure.

\begin{figure}
    \centering
    \includegraphics[scale=1.2]{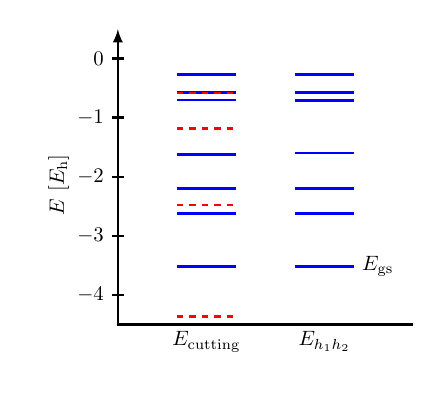}
    \caption{%
      Illustration of the qualitative failure of the cutting projector and the robustness of the newly developed \honehtwo\ projector for the ultra-relativistic $\alpha = 1/2.5$ fine-structure constant value (instead of the CODATA18 $\alpha_0=1/137.035999084$). The energy levels correspond to the Dirac--Coulomb Hamiltonian and the 
      helium atom, the physical ground state is labelled with $E_\text{gs}$. 
      The threshold energy of the cutting projector was $E_\text{th}=-5.1$.
    \label{fig:UltraRelHe}
    }
\end{figure}

Figure~\ref{fig:UltraRelHe} shows an example computation for the helium atom (using $N=100$ ECG functions) corresponding to $\alpha = 1/2.5$, which models an ultra-relativistic situation. 
In this case, we obtained an incorrect state (from the discrete representation of the BR continuum) as the lowest-energy state of the no-pair (cutting) Dirac--Coulomb computation. We can observe in Fig.~\ref{fig:UltraRelHe} that the physical ground-state energy also appears in the energy list, but it is the first excited state. Regarding the higher-energy states, there is (mostly) good agreement between the \honehtwo\ and the cutting energy spectrum, but the cutting energy list contains a few additional (spurious) states, which are attributed to the unphysical BR pollution of the cutting projector space.

Finally, as a technical side remark, our current \honehtwo\ implementation is numerically more sensitive than the cutting projection approach, and all computations in Table~\ref{tab:refdata} required quadruple precision arithmetic, while double precision was sufficient for the cutting projector computations.

\section{Summary, outlook and conclusions \label{ch:summary}}
A consecutive one-particle projection scheme has been proposed for explicitly correlated relativistic computations. All computational steps are carried out over the explicitly correlated, two-particle basis space with the intermediate matrix representation of inherently one-particle operators (\emph{e.g.,} $h_1$) constructed over the entire (permutationally antisymmetric and symmetric) Hilbert space.
As a first application of the one-particle projection scheme, the no-pair relativistic energy of two-electron atomic and molecular systems was computed (within the Born--Oppenheimer approximation), and it was found to be in excellent numerical agreement with the simple energy-cutting projector results within the estimated basis convergence error. 
The $\alpha$ fine-structure constant dependence of the cutting-projector no-pair energies was formerly demonstrated \cite{JeFeMa22,FeJeMa22b,JeMa23,FeMa23} to be in excellent numerical agreement with the relevant, high-precision non-relativistic QED values, used as the current theory benchmark for precision spectroscopy. 

The one-particle projection scheme is expected to perform similarly well for medium-to-high-$Z$ systems, and also in the infinite basis set limit, for which the simple energy cutting fails (for fundamental reasons) and the complex coordinate rotation projector was found to be inefficient. 

Furthermore, the one-particle projection scheme can be extended to construct not only the positive-energy ($++$), but also the negative-energy ($--$) as well as $+-$ and $-+$ subspaces (separately), which is a prerequisite for the evaluation of quantum electrodynamics corrections to the correlated, no-pair energy \cite{MaFeJeMa23,MaMa24,NoMaMa24}.

\begin{acknowledgements}
\noindent Financial support of the European Research Council through a Starting Grant (No.~851421) is gratefully acknowledged.
\end{acknowledgements}


\begin{thebibliography}{72}%
\makeatletter
\providecommand \@ifxundefined [1]{%
 \@ifx{#1\undefined}
}%
\providecommand \@ifnum [1]{%
 \ifnum #1\expandafter \@firstoftwo
 \else \expandafter \@secondoftwo
 \fi
}%
\providecommand \@ifx [1]{%
 \ifx #1\expandafter \@firstoftwo
 \else \expandafter \@secondoftwo
 \fi
}%
\providecommand \natexlab [1]{#1}%
\providecommand \enquote  [1]{``#1''}%
\providecommand \bibnamefont  [1]{#1}%
\providecommand \bibfnamefont [1]{#1}%
\providecommand \citenamefont [1]{#1}%
\providecommand \href@noop [0]{\@secondoftwo}%
\providecommand \href [0]{\begingroup \@sanitize@url \@href}%
\providecommand \@href[1]{\@@startlink{#1}\@@href}%
\providecommand \@@href[1]{\endgroup#1\@@endlink}%
\providecommand \@sanitize@url [0]{\catcode `\\12\catcode `\$12\catcode
  `\&12\catcode `\#12\catcode `\^12\catcode `\_12\catcode `\%12\relax}%
\providecommand \@@startlink[1]{}%
\providecommand \@@endlink[0]{}%
\providecommand \url  [0]{\begingroup\@sanitize@url \@url }%
\providecommand \@url [1]{\endgroup\@href {#1}{\urlprefix }}%
\providecommand \urlprefix  [0]{URL }%
\providecommand \Eprint [0]{\href }%
\providecommand \doibase [0]{http://dx.doi.org/}%
\providecommand \selectlanguage [0]{\@gobble}%
\providecommand \bibinfo  [0]{\@secondoftwo}%
\providecommand \bibfield  [0]{\@secondoftwo}%
\providecommand \translation [1]{[#1]}%
\providecommand \BibitemOpen [0]{}%
\providecommand \bibitemStop [0]{}%
\providecommand \bibitemNoStop [0]{.\EOS\space}%
\providecommand \EOS [0]{\spacefactor3000\relax}%
\providecommand \BibitemShut  [1]{\csname bibitem#1\endcsname}%
\let\auto@bib@innerbib\@empty
\bibitem [{\citenamefont {Beyer}\ \emph {et~al.}(2019)\citenamefont {Beyer},
  \citenamefont {H\"olsch}, \citenamefont {Hussels}, \citenamefont {Cheng},
  \citenamefont {Salumbides}, \citenamefont {Eikema}, \citenamefont {Ubachs},
  \citenamefont {Jungen},\ and\ \citenamefont {Merkt}}]{BeHoHuChSaEiUbJuMe19}%
  \BibitemOpen
  \bibfield  {author} {\bibinfo {author} {\bibfnamefont {M.}~\bibnamefont
  {Beyer}}, \bibinfo {author} {\bibfnamefont {N.}~\bibnamefont {H\"olsch}},
  \bibinfo {author} {\bibfnamefont {J.}~\bibnamefont {Hussels}}, \bibinfo
  {author} {\bibfnamefont {C.-F.}\ \bibnamefont {Cheng}}, \bibinfo {author}
  {\bibfnamefont {E.~J.}\ \bibnamefont {Salumbides}}, \bibinfo {author}
  {\bibfnamefont {K.~S.~E.}\ \bibnamefont {Eikema}}, \bibinfo {author}
  {\bibfnamefont {W.}~\bibnamefont {Ubachs}}, \bibinfo {author} {\bibfnamefont
  {C.}~\bibnamefont {Jungen}}, \ and\ \bibinfo {author} {\bibfnamefont
  {F.}~\bibnamefont {Merkt}},\ }\href {\doibase 10.1103/PhysRevLett.123.163002}
  {\bibfield  {journal} {\bibinfo  {journal} {Phys. Rev. Lett.}\ }\textbf
  {\bibinfo {volume} {123}},\ \bibinfo {pages} {163002} (\bibinfo {year}
  {2019})}\BibitemShut {NoStop}%
\bibitem [{\citenamefont {Semeria}\ \emph {et~al.}(2020)\citenamefont
  {Semeria}, \citenamefont {Jansen}, \citenamefont {Camenisch}, \citenamefont
  {Mellini}, \citenamefont {Schmutz},\ and\ \citenamefont
  {Merkt}}]{SeJaCaMeScMe20}%
  \BibitemOpen
  \bibfield  {author} {\bibinfo {author} {\bibfnamefont {L.}~\bibnamefont
  {Semeria}}, \bibinfo {author} {\bibfnamefont {P.}~\bibnamefont {Jansen}},
  \bibinfo {author} {\bibfnamefont {G.-M.}\ \bibnamefont {Camenisch}}, \bibinfo
  {author} {\bibfnamefont {F.}~\bibnamefont {Mellini}}, \bibinfo {author}
  {\bibfnamefont {H.}~\bibnamefont {Schmutz}}, \ and\ \bibinfo {author}
  {\bibfnamefont {F.}~\bibnamefont {Merkt}},\ }\href {\doibase
  10.1103/PhysRevLett.124.213001} {\bibfield  {journal} {\bibinfo  {journal}
  {Phys. Rev. Lett.}\ }\textbf {\bibinfo {volume} {124}},\ \bibinfo {pages}
  {213001} (\bibinfo {year} {2020})}\BibitemShut {NoStop}%
\bibitem [{\citenamefont {Clausen}\ \emph {et~al.}(2021)\citenamefont
  {Clausen}, \citenamefont {Jansen}, \citenamefont {Scheidegger}, \citenamefont
  {Agner}, \citenamefont {Schmutz},\ and\ \citenamefont
  {Merkt}}]{ClJaScAgScMe21}%
  \BibitemOpen
  \bibfield  {author} {\bibinfo {author} {\bibfnamefont {G.}~\bibnamefont
  {Clausen}}, \bibinfo {author} {\bibfnamefont {P.}~\bibnamefont {Jansen}},
  \bibinfo {author} {\bibfnamefont {S.}~\bibnamefont {Scheidegger}}, \bibinfo
  {author} {\bibfnamefont {J.~A.}\ \bibnamefont {Agner}}, \bibinfo {author}
  {\bibfnamefont {H.}~\bibnamefont {Schmutz}}, \ and\ \bibinfo {author}
  {\bibfnamefont {F.}~\bibnamefont {Merkt}},\ }\href {\doibase
  10.1103/PhysRevLett.127.093001} {\bibfield  {journal} {\bibinfo  {journal}
  {Phys. Rev. Lett.}\ }\textbf {\bibinfo {volume} {127}},\ \bibinfo {pages}
  {093001} (\bibinfo {year} {2021})}\BibitemShut {NoStop}%
\bibitem [{\citenamefont {Gurung}\ \emph {et~al.}(2021)\citenamefont {Gurung},
  \citenamefont {Babij}, \citenamefont {P\'erez-R\'{\i}os}, \citenamefont
  {Hogan},\ and\ \citenamefont {Cassidy}}]{GuBaPRHoCa21}%
  \BibitemOpen
  \bibfield  {author} {\bibinfo {author} {\bibfnamefont {L.}~\bibnamefont
  {Gurung}}, \bibinfo {author} {\bibfnamefont {T.~J.}\ \bibnamefont {Babij}},
  \bibinfo {author} {\bibfnamefont {J.}~\bibnamefont {P\'erez-R\'{\i}os}},
  \bibinfo {author} {\bibfnamefont {S.~D.}\ \bibnamefont {Hogan}}, \ and\
  \bibinfo {author} {\bibfnamefont {D.~B.}\ \bibnamefont {Cassidy}},\ }\href
  {\doibase 10.1103/PhysRevA.103.042805} {\bibfield  {journal} {\bibinfo
  {journal} {Phys. Rev. A}\ }\textbf {\bibinfo {volume} {103}},\ \bibinfo
  {pages} {042805} (\bibinfo {year} {2021})}\BibitemShut {NoStop}%
\bibitem [{\citenamefont {Sheldon}\ \emph {et~al.}(2023)\citenamefont
  {Sheldon}, \citenamefont {Babij}, \citenamefont {Reeder}, \citenamefont
  {Hogan},\ and\ \citenamefont {Cassidy}}]{ShBaReHoCa23}%
  \BibitemOpen
  \bibfield  {author} {\bibinfo {author} {\bibfnamefont {R.~E.}\ \bibnamefont
  {Sheldon}}, \bibinfo {author} {\bibfnamefont {T.~J.}\ \bibnamefont {Babij}},
  \bibinfo {author} {\bibfnamefont {S.~H.}\ \bibnamefont {Reeder}}, \bibinfo
  {author} {\bibfnamefont {S.~D.}\ \bibnamefont {Hogan}}, \ and\ \bibinfo
  {author} {\bibfnamefont {D.~B.}\ \bibnamefont {Cassidy}},\ }\href {\doibase
  10.1103/PhysRevLett.131.043001} {\bibfield  {journal} {\bibinfo  {journal}
  {Phys. Rev. Lett.}\ }\textbf {\bibinfo {volume} {131}},\ \bibinfo {pages}
  {043001} (\bibinfo {year} {2023})}\BibitemShut {NoStop}%
\bibitem [{\citenamefont {Clausen}\ \emph {et~al.}(2023)\citenamefont
  {Clausen}, \citenamefont {Scheidegger}, \citenamefont {Agner}, \citenamefont
  {Schmutz},\ and\ \citenamefont {Merkt}}]{ClScAgScMe23}%
  \BibitemOpen
  \bibfield  {author} {\bibinfo {author} {\bibfnamefont {G.}~\bibnamefont
  {Clausen}}, \bibinfo {author} {\bibfnamefont {S.}~\bibnamefont
  {Scheidegger}}, \bibinfo {author} {\bibfnamefont {J.~A.}\ \bibnamefont
  {Agner}}, \bibinfo {author} {\bibfnamefont {H.}~\bibnamefont {Schmutz}}, \
  and\ \bibinfo {author} {\bibfnamefont {F.}~\bibnamefont {Merkt}},\ }\href
  {\doibase 10.1103/PhysRevLett.131.103001} {\bibfield  {journal} {\bibinfo
  {journal} {Phys. Rev. Lett.}\ }\textbf {\bibinfo {volume} {131}},\ \bibinfo
  {pages} {103001} (\bibinfo {year} {2023})}\BibitemShut {NoStop}%
\bibitem [{\citenamefont {Germann}\ \emph {et~al.}(2021)\citenamefont
  {Germann}, \citenamefont {Patra}, \citenamefont {Karr}, \citenamefont
  {Hilico}, \citenamefont {Korobov}, \citenamefont {Salumbides}, \citenamefont
  {Eikema}, \citenamefont {Ubachs},\ and\ \citenamefont {Koelemeij}}]{test21}%
  \BibitemOpen
  \bibfield  {author} {\bibinfo {author} {\bibfnamefont {M.}~\bibnamefont
  {Germann}}, \bibinfo {author} {\bibfnamefont {S.}~\bibnamefont {Patra}},
  \bibinfo {author} {\bibfnamefont {J.-P.}\ \bibnamefont {Karr}}, \bibinfo
  {author} {\bibfnamefont {L.}~\bibnamefont {Hilico}}, \bibinfo {author}
  {\bibfnamefont {V.~I.}\ \bibnamefont {Korobov}}, \bibinfo {author}
  {\bibfnamefont {E.~J.}\ \bibnamefont {Salumbides}}, \bibinfo {author}
  {\bibfnamefont {K.~S.~E.}\ \bibnamefont {Eikema}}, \bibinfo {author}
  {\bibfnamefont {W.}~\bibnamefont {Ubachs}}, \ and\ \bibinfo {author}
  {\bibfnamefont {J.~C.~J.}\ \bibnamefont {Koelemeij}},\ }\href {\doibase
  10.1103/PhysRevResearch.3.L022028} {\bibfield  {journal} {\bibinfo  {journal}
  {Phys. Rev. Res.}\ }\textbf {\bibinfo {volume} {3}},\ \bibinfo {pages}
  {L022028} (\bibinfo {year} {2021})}\BibitemShut {NoStop}%
\bibitem [{\citenamefont {Alighanbari}\ \emph {et~al.}(2020)\citenamefont
  {Alighanbari}, \citenamefont {Giri}, \citenamefont {Constantin},
  \citenamefont {Korobov},\ and\ \citenamefont {Schiller}}]{AlGiCoKoSc20}%
  \BibitemOpen
  \bibfield  {author} {\bibinfo {author} {\bibfnamefont {S.}~\bibnamefont
  {Alighanbari}}, \bibinfo {author} {\bibfnamefont {G.~S.}\ \bibnamefont
  {Giri}}, \bibinfo {author} {\bibfnamefont {F.~L.}\ \bibnamefont
  {Constantin}}, \bibinfo {author} {\bibfnamefont {V.~I.}\ \bibnamefont
  {Korobov}}, \ and\ \bibinfo {author} {\bibfnamefont {S.}~\bibnamefont
  {Schiller}},\ }\href {\doibase 10.1038/s41586-020-2261-5} {\bibfield
  {journal} {\bibinfo  {journal} {Nature}\ }\textbf {\bibinfo {volume} {581}},\
  \bibinfo {pages} {152} (\bibinfo {year} {2020})}\BibitemShut {NoStop}%
\bibitem [{\citenamefont {Patra}\ \emph {et~al.}(2020)\citenamefont {Patra},
  \citenamefont {Germann}, \citenamefont {Karr}, \citenamefont {Haidar},
  \citenamefont {Hilico}, \citenamefont {Korobov}, \citenamefont {Cozijn},
  \citenamefont {Eikema}, \citenamefont {Ubachs},\ and\ \citenamefont
  {Koelemeij}}]{sci2020}%
  \BibitemOpen
  \bibfield  {author} {\bibinfo {author} {\bibfnamefont {S.}~\bibnamefont
  {Patra}}, \bibinfo {author} {\bibfnamefont {M.}~\bibnamefont {Germann}},
  \bibinfo {author} {\bibfnamefont {J.-P.}\ \bibnamefont {Karr}}, \bibinfo
  {author} {\bibfnamefont {M.}~\bibnamefont {Haidar}}, \bibinfo {author}
  {\bibfnamefont {L.}~\bibnamefont {Hilico}}, \bibinfo {author} {\bibfnamefont
  {V.~I.}\ \bibnamefont {Korobov}}, \bibinfo {author} {\bibfnamefont
  {F.~M.~J.}\ \bibnamefont {Cozijn}}, \bibinfo {author} {\bibfnamefont
  {K.~S.~E.}\ \bibnamefont {Eikema}}, \bibinfo {author} {\bibfnamefont
  {W.}~\bibnamefont {Ubachs}}, \ and\ \bibinfo {author} {\bibfnamefont
  {J.~C.~J.}\ \bibnamefont {Koelemeij}},\ }\href {\doibase
  10.1126/science.aba0453} {\bibfield  {journal} {\bibinfo  {journal}
  {Science}\ }\textbf {\bibinfo {volume} {369}},\ \bibinfo {pages} {1238}
  (\bibinfo {year} {2020})}\BibitemShut {NoStop}%
\bibitem [{\citenamefont {Yelkhovsky}(2001)}]{Ye01}%
  \BibitemOpen
  \bibfield  {author} {\bibinfo {author} {\bibfnamefont {A.}~\bibnamefont
  {Yelkhovsky}},\ }\href {\doibase 10.1103/PhysRevA.64.062104} {\bibfield
  {journal} {\bibinfo  {journal} {Phys. Rev. A}\ }\textbf {\bibinfo {volume}
  {64}},\ \bibinfo {pages} {062104} (\bibinfo {year} {2001})}\BibitemShut
  {NoStop}%
\bibitem [{\citenamefont {Korobov}\ and\ \citenamefont
  {Yelkhovsky}(2001)}]{KoYe01}%
  \BibitemOpen
  \bibfield  {author} {\bibinfo {author} {\bibfnamefont {V.}~\bibnamefont
  {Korobov}}\ and\ \bibinfo {author} {\bibfnamefont {A.}~\bibnamefont
  {Yelkhovsky}},\ }\href {\doibase 10.1103/PhysRevLett.87.193003} {\bibfield
  {journal} {\bibinfo  {journal} {Phys. Rev. Lett.}\ }\textbf {\bibinfo
  {volume} {87}},\ \bibinfo {pages} {193003} (\bibinfo {year}
  {2001})}\BibitemShut {NoStop}%
\bibitem [{\citenamefont {Pachucki}(2006)}]{Pa06}%
  \BibitemOpen
  \bibfield  {author} {\bibinfo {author} {\bibfnamefont {K.}~\bibnamefont
  {Pachucki}},\ }\href {\doibase 10.1103/PhysRevA.74.022512} {\bibfield
  {journal} {\bibinfo  {journal} {Phys. Rev. A}\ }\textbf {\bibinfo {volume}
  {74}},\ \bibinfo {pages} {022512} (\bibinfo {year} {2006})}\BibitemShut
  {NoStop}%
\bibitem [{\citenamefont {Korobov}\ and\ \citenamefont
  {Tsogbayar}(2007)}]{KoTs07}%
  \BibitemOpen
  \bibfield  {author} {\bibinfo {author} {\bibfnamefont {V.~I.}\ \bibnamefont
  {Korobov}}\ and\ \bibinfo {author} {\bibfnamefont {T.}~\bibnamefont
  {Tsogbayar}},\ }\href {\doibase 10.1088/0953-4075/40/13/011} {\bibfield
  {journal} {\bibinfo  {journal} {J. Phys. B}\ }\textbf {\bibinfo {volume}
  {40}},\ \bibinfo {pages} {2661} (\bibinfo {year} {2007})}\BibitemShut
  {NoStop}%
\bibitem [{\citenamefont {Korobov}\ \emph {et~al.}(2013)\citenamefont
  {Korobov}, \citenamefont {Hilico},\ and\ \citenamefont {Karr}}]{KoHiKa13}%
  \BibitemOpen
  \bibfield  {author} {\bibinfo {author} {\bibfnamefont {V.~I.}\ \bibnamefont
  {Korobov}}, \bibinfo {author} {\bibfnamefont {L.}~\bibnamefont {Hilico}}, \
  and\ \bibinfo {author} {\bibfnamefont {J.-P.}\ \bibnamefont {Karr}},\ }\href
  {\doibase 10.1103/PhysRevA.87.062506} {\bibfield  {journal} {\bibinfo
  {journal} {Phys. Rev. A}\ }\textbf {\bibinfo {volume} {87}},\ \bibinfo
  {pages} {062506} (\bibinfo {year} {2013})}\BibitemShut {NoStop}%
\bibitem [{\citenamefont {Korobov}\ \emph {et~al.}(2014)\citenamefont
  {Korobov}, \citenamefont {Hilico},\ and\ \citenamefont {Karr}}]{KoHiKa14}%
  \BibitemOpen
  \bibfield  {author} {\bibinfo {author} {\bibfnamefont {V.~I.}\ \bibnamefont
  {Korobov}}, \bibinfo {author} {\bibfnamefont {L.}~\bibnamefont {Hilico}}, \
  and\ \bibinfo {author} {\bibfnamefont {J.-P.}\ \bibnamefont {Karr}},\ }\href
  {\doibase 10.1103/PhysRevLett.112.103003} {\bibfield  {journal} {\bibinfo
  {journal} {Phys. Rev. Lett.}\ }\textbf {\bibinfo {volume} {112}},\ \bibinfo
  {pages} {103003} (\bibinfo {year} {2014})}\BibitemShut {NoStop}%
\bibitem [{\citenamefont {Patk\'o\ifmmode~\check{s}\else \v{s}\fi{}}\ \emph
  {et~al.}(2019)\citenamefont {Patk\'o\ifmmode~\check{s}\else \v{s}\fi{}},
  \citenamefont {Yerokhin},\ and\ \citenamefont {Pachucki}}]{PaYePa19}%
  \BibitemOpen
  \bibfield  {author} {\bibinfo {author} {\bibfnamefont {V.}~\bibnamefont
  {Patk\'o\ifmmode~\check{s}\else \v{s}\fi{}}}, \bibinfo {author}
  {\bibfnamefont {V.~A.}\ \bibnamefont {Yerokhin}}, \ and\ \bibinfo {author}
  {\bibfnamefont {K.}~\bibnamefont {Pachucki}},\ }\href {\doibase
  10.1103/PhysRevA.100.042510} {\bibfield  {journal} {\bibinfo  {journal}
  {Phys. Rev. A}\ }\textbf {\bibinfo {volume} {100}},\ \bibinfo {pages}
  {042510} (\bibinfo {year} {2019})}\BibitemShut {NoStop}%
\bibitem [{\citenamefont {Patk\'o\ifmmode~\check{s}\else \v{s}\fi{}}\ \emph
  {et~al.}(2020)\citenamefont {Patk\'o\ifmmode~\check{s}\else \v{s}\fi{}},
  \citenamefont {Yerokhin},\ and\ \citenamefont {Pachucki}}]{PaYePa20}%
  \BibitemOpen
  \bibfield  {author} {\bibinfo {author} {\bibfnamefont {V.}~\bibnamefont
  {Patk\'o\ifmmode~\check{s}\else \v{s}\fi{}}}, \bibinfo {author}
  {\bibfnamefont {V.~A.}\ \bibnamefont {Yerokhin}}, \ and\ \bibinfo {author}
  {\bibfnamefont {K.}~\bibnamefont {Pachucki}},\ }\href {\doibase
  10.1103/PhysRevA.101.062516} {\bibfield  {journal} {\bibinfo  {journal}
  {Phys. Rev. A}\ }\textbf {\bibinfo {volume} {101}},\ \bibinfo {pages}
  {062516} (\bibinfo {year} {2020})}\BibitemShut {NoStop}%
\bibitem [{\citenamefont {Patk\'o\ifmmode~\check{s}\else \v{s}\fi{}}\ \emph
  {et~al.}(2021)\citenamefont {Patk\'o\ifmmode~\check{s}\else \v{s}\fi{}},
  \citenamefont {Yerokhin},\ and\ \citenamefont {Pachucki}}]{PaYeVlPa21}%
  \BibitemOpen
  \bibfield  {author} {\bibinfo {author} {\bibfnamefont {V.}~\bibnamefont
  {Patk\'o\ifmmode~\check{s}\else \v{s}\fi{}}}, \bibinfo {author}
  {\bibfnamefont {V.~A.}\ \bibnamefont {Yerokhin}}, \ and\ \bibinfo {author}
  {\bibfnamefont {K.}~\bibnamefont {Pachucki}},\ }\href {\doibase
  10.1103/PhysRevA.103.042809} {\bibfield  {journal} {\bibinfo  {journal}
  {Phys. Rev. A}\ }\textbf {\bibinfo {volume} {103}},\ \bibinfo {pages}
  {042809} (\bibinfo {year} {2021})}\BibitemShut {NoStop}%
\bibitem [{\citenamefont {Yerokhin}\ \emph {et~al.}(2022)\citenamefont
  {Yerokhin}, \citenamefont {Patk\'o\ifmmode~\check{s}\else \v{s}\fi{}},\ and\
  \citenamefont {Pachucki}}]{YePaPa22}%
  \BibitemOpen
  \bibfield  {author} {\bibinfo {author} {\bibfnamefont {V.~A.}\ \bibnamefont
  {Yerokhin}}, \bibinfo {author} {\bibfnamefont {V.~c.~v.}\ \bibnamefont
  {Patk\'o\ifmmode~\check{s}\else \v{s}\fi{}}}, \ and\ \bibinfo {author}
  {\bibfnamefont {K.}~\bibnamefont {Pachucki}},\ }\href {\doibase
  10.1103/PhysRevA.106.022815} {\bibfield  {journal} {\bibinfo  {journal}
  {Phys. Rev. A}\ }\textbf {\bibinfo {volume} {106}},\ \bibinfo {pages}
  {022815} (\bibinfo {year} {2022})}\BibitemShut {NoStop}%
\bibitem [{\citenamefont {Nogueira}\ and\ \citenamefont {Karr}(2023)}]{NoKa23}%
  \BibitemOpen
  \bibfield  {author} {\bibinfo {author} {\bibfnamefont {H.~D.}\ \bibnamefont
  {Nogueira}}\ and\ \bibinfo {author} {\bibfnamefont {J.-P.}\ \bibnamefont
  {Karr}},\ }\href {\doibase 10.1103/PhysRevA.107.042817} {\bibfield  {journal}
  {\bibinfo  {journal} {Phys. Rev. A}\ }\textbf {\bibinfo {volume} {107}},\
  \bibinfo {pages} {042817} (\bibinfo {year} {2023})}\BibitemShut {NoStop}%
\bibitem [{\citenamefont {Kullie}\ and\ \citenamefont
  {Schiller}(2022)}]{KuSc22}%
  \BibitemOpen
  \bibfield  {author} {\bibinfo {author} {\bibfnamefont {O.}~\bibnamefont
  {Kullie}}\ and\ \bibinfo {author} {\bibfnamefont {S.}~\bibnamefont
  {Schiller}},\ }\href {\doibase 10.1103/PhysRevA.105.052801} {\bibfield
  {journal} {\bibinfo  {journal} {Phys. Rev. A}\ }\textbf {\bibinfo {volume}
  {105}},\ \bibinfo {pages} {052801} (\bibinfo {year} {2022})}\BibitemShut
  {NoStop}%
\bibitem [{\citenamefont {M\'atyus}\ \emph {et~al.}(2023)\citenamefont
  {M\'atyus}, \citenamefont {Ferenc}, \citenamefont {Jeszenszki},\ and\
  \citenamefont {Marg\'ocsy}}]{MaFeJeMa23}%
  \BibitemOpen
  \bibfield  {author} {\bibinfo {author} {\bibfnamefont {E.}~\bibnamefont
  {M\'atyus}}, \bibinfo {author} {\bibfnamefont {D.}~\bibnamefont {Ferenc}},
  \bibinfo {author} {\bibfnamefont {P.}~\bibnamefont {Jeszenszki}}, \ and\
  \bibinfo {author} {\bibfnamefont {A.}~\bibnamefont {Marg\'ocsy}},\ }\href
  {\doibase https://doi.org/10.1021/acsphyschemau.2c00062} {\bibfield
  {journal} {\bibinfo  {journal} {ACS Phys. Chem Au}\ }\textbf {\bibinfo
  {volume} {3}},\ \bibinfo {pages} {222} (\bibinfo {year} {2023})}\BibitemShut
  {NoStop}%
\bibitem [{\citenamefont {Margócsy}\ and\ \citenamefont
  {Mátyus}(2024)}]{MaMa24}%
  \BibitemOpen
  \bibfield  {author} {\bibinfo {author} {\bibfnamefont {A.}~\bibnamefont
  {Margócsy}}\ and\ \bibinfo {author} {\bibfnamefont {E.}~\bibnamefont
  {Mátyus}},\ }\href {\doibase 10.1063/5.0193250} {\bibfield  {journal}
  {\bibinfo  {journal} {J. Chem. Phys.}\ }\textbf {\bibinfo {volume} {160}},\
  \bibinfo {pages} {204103} (\bibinfo {year} {2024})}\BibitemShut {NoStop}%
\bibitem [{\citenamefont {Nonn}\ \emph {et~al.}(2024)\citenamefont {Nonn},
  \citenamefont {Margócsy},\ and\ \citenamefont {Mátyus}}]{NoMaMa24}%
  \BibitemOpen
  \bibfield  {author} {\bibinfo {author} {\bibfnamefont {A.}~\bibnamefont
  {Nonn}}, \bibinfo {author} {\bibfnamefont {A.}~\bibnamefont {Margócsy}}, \
  and\ \bibinfo {author} {\bibfnamefont {E.}~\bibnamefont {Mátyus}},\ }\href
  {\doibase 10.1021/acs.jctc.4c00128MaMa} {\bibfield  {journal} {\bibinfo
  {journal} {J. Chem. Theory Comput.}\ } (\bibinfo {year} {2024}),\
  10.1021/acs.jctc.4c00128MaMa}\BibitemShut {NoStop}%
\bibitem [{\citenamefont {Salpeter}\ and\ \citenamefont
  {Bethe}(1951)}]{SaBe51}%
  \BibitemOpen
  \bibfield  {author} {\bibinfo {author} {\bibfnamefont {E.~E.}\ \bibnamefont
  {Salpeter}}\ and\ \bibinfo {author} {\bibfnamefont {H.~A.}\ \bibnamefont
  {Bethe}},\ }\href {\doibase 10.1103/PhysRev.84.1232} {\bibfield  {journal}
  {\bibinfo  {journal} {Phys. Rev.}\ }\textbf {\bibinfo {volume} {84}},\
  \bibinfo {pages} {1232} (\bibinfo {year} {1951})}\BibitemShut {NoStop}%
\bibitem [{\citenamefont {Salpeter}(1952)}]{Sa52}%
  \BibitemOpen
  \bibfield  {author} {\bibinfo {author} {\bibfnamefont {E.~E.}\ \bibnamefont
  {Salpeter}},\ }\href {\doibase 10.1103/PhysRev.87.328} {\bibfield  {journal}
  {\bibinfo  {journal} {Phys. Rev.}\ }\textbf {\bibinfo {volume} {87}},\
  \bibinfo {pages} {328} (\bibinfo {year} {1952})}\BibitemShut {NoStop}%
\bibitem [{\citenamefont {Sucher}(1958)}]{sucherPhD1958}%
  \BibitemOpen
  \bibfield  {author} {\bibinfo {author} {\bibfnamefont {J.}~\bibnamefont
  {Sucher}},\ }\href@noop {} {\enquote {\bibinfo {title} {Energy levels of the
  two-electron atom, to order $\alpha^3${Rydberg} ({C}olumbia {U}niversity)},}\
  }\bibinfo {howpublished} {Ph.D. Thesis} (\bibinfo {year} {1958})\BibitemShut
  {NoStop}%
\bibitem [{\citenamefont {Araki}(1957)}]{araki57}%
  \BibitemOpen
  \bibfield  {author} {\bibinfo {author} {\bibfnamefont {H.}~\bibnamefont
  {Araki}},\ }\href {\doibase 10.1143/PTP.17.619} {\bibfield  {journal}
  {\bibinfo  {journal} {Prog. of Theor. Phys.}\ }\textbf {\bibinfo {volume}
  {17}},\ \bibinfo {pages} {619} (\bibinfo {year} {1957})}\BibitemShut
  {NoStop}%
\bibitem [{\citenamefont {Bylicki}\ \emph {et~al.}(2008)\citenamefont
  {Bylicki}, \citenamefont {Pestka},\ and\ \citenamefont
  {Karwowski}}]{ByPeKa08}%
  \BibitemOpen
  \bibfield  {author} {\bibinfo {author} {\bibfnamefont {M.}~\bibnamefont
  {Bylicki}}, \bibinfo {author} {\bibfnamefont {G.}~\bibnamefont {Pestka}}, \
  and\ \bibinfo {author} {\bibfnamefont {J.}~\bibnamefont {Karwowski}},\ }\href
  {\doibase 10.1103/PhysRevA.77.044501} {\bibfield  {journal} {\bibinfo
  {journal} {Phys. Rev. A}\ }\textbf {\bibinfo {volume} {77}},\ \bibinfo
  {pages} {044501} (\bibinfo {year} {2008})}\BibitemShut {NoStop}%
\bibitem [{\citenamefont {Pestka}\ \emph {et~al.}(2006)\citenamefont {Pestka},
  \citenamefont {Bylicki},\ and\ \citenamefont {Karwowski}}]{PeByKa06}%
  \BibitemOpen
  \bibfield  {author} {\bibinfo {author} {\bibfnamefont {G.}~\bibnamefont
  {Pestka}}, \bibinfo {author} {\bibfnamefont {M.}~\bibnamefont {Bylicki}}, \
  and\ \bibinfo {author} {\bibfnamefont {J.}~\bibnamefont {Karwowski}},\ }\href
  {\doibase 10.1088/0953-4075/39/14/006} {\bibfield  {journal} {\bibinfo
  {journal} {J. Phys. B}\ }\textbf {\bibinfo {volume} {39}},\ \bibinfo {pages}
  {2979} (\bibinfo {year} {2006})}\BibitemShut {NoStop}%
\bibitem [{\citenamefont {Pestka}\ \emph {et~al.}(2007)\citenamefont {Pestka},
  \citenamefont {Bylicki},\ and\ \citenamefont {Karwowski}}]{PeByKa07}%
  \BibitemOpen
  \bibfield  {author} {\bibinfo {author} {\bibfnamefont {G.}~\bibnamefont
  {Pestka}}, \bibinfo {author} {\bibfnamefont {M.}~\bibnamefont {Bylicki}}, \
  and\ \bibinfo {author} {\bibfnamefont {J.}~\bibnamefont {Karwowski}},\ }\href
  {\doibase 10.1088/0953-4075/40/12/003} {\bibfield  {journal} {\bibinfo
  {journal} {J. Phys. B}\ }\textbf {\bibinfo {volume} {40}},\ \bibinfo {pages}
  {2249} (\bibinfo {year} {2007})}\BibitemShut {NoStop}%
\bibitem [{\citenamefont {Pestka}\ \emph {et~al.}(2012)\citenamefont {Pestka},
  \citenamefont {Bylicki},\ and\ \citenamefont {Karwowski}}]{PeByKa12}%
  \BibitemOpen
  \bibfield  {author} {\bibinfo {author} {\bibfnamefont {G.}~\bibnamefont
  {Pestka}}, \bibinfo {author} {\bibfnamefont {M.}~\bibnamefont {Bylicki}}, \
  and\ \bibinfo {author} {\bibfnamefont {J.}~\bibnamefont {Karwowski}},\ }\href
  {\doibase 10.1007/s10910-011-9823-6} {\bibfield  {journal} {\bibinfo
  {journal} {J. Math. Chem.}\ }\textbf {\bibinfo {volume} {50}},\ \bibinfo
  {pages} {510} (\bibinfo {year} {2012})}\BibitemShut {NoStop}%
\bibitem [{\citenamefont {Karwowski}(2017)}]{Ka17}%
  \BibitemOpen
  \bibfield  {author} {\bibinfo {author} {\bibfnamefont {J.}~\bibnamefont
  {Karwowski}},\ }\enquote {\bibinfo {title} {Dirac operator and its
  properties},}\ \ (\bibinfo  {publisher} {Springer},\ \bibinfo {address}
  {Berlin, Heidelberg},\ \bibinfo {year} {2017})\ pp.\ \bibinfo {pages}
  {3--49}\BibitemShut {NoStop}%
\bibitem [{\citenamefont {Jeszenszki}\ \emph {et~al.}(2021)\citenamefont
  {Jeszenszki}, \citenamefont {Ferenc},\ and\ \citenamefont
  {M\'atyus}}]{JeFeMa21}%
  \BibitemOpen
  \bibfield  {author} {\bibinfo {author} {\bibfnamefont {P.}~\bibnamefont
  {Jeszenszki}}, \bibinfo {author} {\bibfnamefont {D.}~\bibnamefont {Ferenc}},
  \ and\ \bibinfo {author} {\bibfnamefont {E.}~\bibnamefont {M\'atyus}},\
  }\href {\doibase 10.1063/5.0051237} {\bibfield  {journal} {\bibinfo
  {journal} {J. Chem. Phys.}\ }\textbf {\bibinfo {volume} {154}},\ \bibinfo
  {pages} {224110} (\bibinfo {year} {2021})}\BibitemShut {NoStop}%
\bibitem [{\citenamefont {Jeszenszki}\ \emph
  {et~al.}(2022{\natexlab{a}})\citenamefont {Jeszenszki}, \citenamefont
  {Ferenc},\ and\ \citenamefont {M\'atyus}}]{JeFeMa22}%
  \BibitemOpen
  \bibfield  {author} {\bibinfo {author} {\bibfnamefont {P.}~\bibnamefont
  {Jeszenszki}}, \bibinfo {author} {\bibfnamefont {D.}~\bibnamefont {Ferenc}},
  \ and\ \bibinfo {author} {\bibfnamefont {E.}~\bibnamefont {M\'atyus}},\
  }\href {\doibase 10.1063/5.0075096} {\bibfield  {journal} {\bibinfo
  {journal} {J. Chem. Phys.}\ }\textbf {\bibinfo {volume} {156}},\ \bibinfo
  {pages} {084111} (\bibinfo {year} {2022}{\natexlab{a}})}\BibitemShut
  {NoStop}%
\bibitem [{\citenamefont {Ferenc}\ \emph
  {et~al.}(2022{\natexlab{a}})\citenamefont {Ferenc}, \citenamefont
  {Jeszenszki},\ and\ \citenamefont {M\'atyus}}]{FeJeMa22}%
  \BibitemOpen
  \bibfield  {author} {\bibinfo {author} {\bibfnamefont {D.}~\bibnamefont
  {Ferenc}}, \bibinfo {author} {\bibfnamefont {P.}~\bibnamefont {Jeszenszki}},
  \ and\ \bibinfo {author} {\bibfnamefont {E.}~\bibnamefont {M\'atyus}},\
  }\href {\doibase 10.1063/5.0075097} {\bibfield  {journal} {\bibinfo
  {journal} {J. Chem. Phys.}\ }\textbf {\bibinfo {volume} {156}},\ \bibinfo
  {pages} {084110} (\bibinfo {year} {2022}{\natexlab{a}})}\BibitemShut
  {NoStop}%
\bibitem [{\citenamefont {Ferenc}\ \emph
  {et~al.}(2022{\natexlab{b}})\citenamefont {Ferenc}, \citenamefont
  {Jeszenszki},\ and\ \citenamefont {Matyus}}]{FeJeMa22b}%
  \BibitemOpen
  \bibfield  {author} {\bibinfo {author} {\bibfnamefont {D.}~\bibnamefont
  {Ferenc}}, \bibinfo {author} {\bibfnamefont {P.}~\bibnamefont {Jeszenszki}},
  \ and\ \bibinfo {author} {\bibfnamefont {E.}~\bibnamefont {Matyus}},\ }\href
  {\doibase 10.1063/5.0105355} {\bibfield  {journal} {\bibinfo  {journal} {J.
  Chem. Phys.}\ }\textbf {\bibinfo {volume} {157}},\ \bibinfo {pages} {094113}
  (\bibinfo {year} {2022}{\natexlab{b}})}\BibitemShut {NoStop}%
\bibitem [{\citenamefont {{J}eszenszki}\ and\ \citenamefont
  {{M}átyus}(2023)}]{JeMa23}%
  \BibitemOpen
  \bibfield  {author} {\bibinfo {author} {\bibfnamefont {P.}~\bibnamefont
  {{J}eszenszki}}\ and\ \bibinfo {author} {\bibfnamefont {E.}~\bibnamefont
  {{M}átyus}},\ }\href {\doibase 10.1063/5.0136360} {\bibfield  {journal}
  {\bibinfo  {journal} {J. Chem. Phys.}\ }\textbf {\bibinfo {volume} {158}},\
  \bibinfo {pages} {054104} (\bibinfo {year} {2023})}\BibitemShut {NoStop}%
\bibitem [{\citenamefont {Ferenc}\ and\ \citenamefont
  {M\'atyus}(2023)}]{FeMa23}%
  \BibitemOpen
  \bibfield  {author} {\bibinfo {author} {\bibfnamefont {D.}~\bibnamefont
  {Ferenc}}\ and\ \bibinfo {author} {\bibfnamefont {E.}~\bibnamefont
  {M\'atyus}},\ }\href {\doibase 10.1103/PhysRevA.107.052803} {\bibfield
  {journal} {\bibinfo  {journal} {Phys. Rev. A}\ }\textbf {\bibinfo {volume}
  {107}},\ \bibinfo {pages} {052803} (\bibinfo {year} {2023})}\BibitemShut
  {NoStop}%
\bibitem [{\citenamefont {Suzuki}\ and\ \citenamefont {Varga}(1998)}]{SuVa98}%
  \BibitemOpen
  \bibfield  {author} {\bibinfo {author} {\bibfnamefont {Y.}~\bibnamefont
  {Suzuki}}\ and\ \bibinfo {author} {\bibfnamefont {K.}~\bibnamefont {Varga}},\
  }\href {\doibase 10.1007/3-540-49541-X} {\emph {\bibinfo {title}
  {{S}tochastic {V}ariational {A}pproach to {Q}uantum{-}{M}echanical
  {F}ew{-}{B}ody {P}roblems}}}\ (\bibinfo  {publisher} {Springer-Verlag},\
  \bibinfo {address} {Berlin, Heidelberg},\ \bibinfo {year} {1998})\BibitemShut
  {NoStop}%
\bibitem [{\citenamefont {Mitroy}\ \emph {et~al.}(2013)\citenamefont {Mitroy},
  \citenamefont {Bubin}, \citenamefont {Horiuchi}, \citenamefont {Suzuki},
  \citenamefont {Adamowicz}, \citenamefont {Cencek}, \citenamefont {Szalewicz},
  \citenamefont {Komasa}, \citenamefont {Blume},\ and\ \citenamefont
  {Varga}}]{rmp13}%
  \BibitemOpen
  \bibfield  {author} {\bibinfo {author} {\bibfnamefont {J.}~\bibnamefont
  {Mitroy}}, \bibinfo {author} {\bibfnamefont {S.}~\bibnamefont {Bubin}},
  \bibinfo {author} {\bibfnamefont {W.}~\bibnamefont {Horiuchi}}, \bibinfo
  {author} {\bibfnamefont {Y.}~\bibnamefont {Suzuki}}, \bibinfo {author}
  {\bibfnamefont {L.}~\bibnamefont {Adamowicz}}, \bibinfo {author}
  {\bibfnamefont {W.}~\bibnamefont {Cencek}}, \bibinfo {author} {\bibfnamefont
  {K.}~\bibnamefont {Szalewicz}}, \bibinfo {author} {\bibfnamefont
  {J.}~\bibnamefont {Komasa}}, \bibinfo {author} {\bibfnamefont
  {D.}~\bibnamefont {Blume}}, \ and\ \bibinfo {author} {\bibfnamefont
  {K.}~\bibnamefont {Varga}},\ }\href@noop {} {\bibfield  {journal} {\bibinfo
  {journal} {Rev. Mod. Phys.}\ }\textbf {\bibinfo {volume} {85}},\ \bibinfo
  {pages} {693} (\bibinfo {year} {2013})}\BibitemShut {NoStop}%
\bibitem [{\citenamefont {M\'atyus}\ and\ \citenamefont
  {Reiher}(2012)}]{MaRe12}%
  \BibitemOpen
  \bibfield  {author} {\bibinfo {author} {\bibfnamefont {E.}~\bibnamefont
  {M\'atyus}}\ and\ \bibinfo {author} {\bibfnamefont {M.}~\bibnamefont
  {Reiher}},\ }\href {\doibase 10.1063/1.4731696} {\bibfield  {journal}
  {\bibinfo  {journal} {J. Chem. Phys.}\ }\textbf {\bibinfo {volume} {137}},\
  \bibinfo {pages} {024104} (\bibinfo {year} {2012})}\BibitemShut {NoStop}%
\bibitem [{\citenamefont {Jeszenszki}\ \emph
  {et~al.}(2022{\natexlab{b}})\citenamefont {Jeszenszki}, \citenamefont
  {Ireland}, \citenamefont {Ferenc},\ and\ \citenamefont
  {Mátyus}}]{JeIrFeMa22}%
  \BibitemOpen
  \bibfield  {author} {\bibinfo {author} {\bibfnamefont {P.}~\bibnamefont
  {Jeszenszki}}, \bibinfo {author} {\bibfnamefont {R.~T.}\ \bibnamefont
  {Ireland}}, \bibinfo {author} {\bibfnamefont {D.}~\bibnamefont {Ferenc}}, \
  and\ \bibinfo {author} {\bibfnamefont {E.}~\bibnamefont {Mátyus}},\ }\href
  {\doibase https://doi.org/10.1002/qua.26819} {\bibfield  {journal} {\bibinfo
  {journal} {Int. J. Quant. Chem.}\ }\textbf {\bibinfo {volume} {122}},\
  \bibinfo {pages} {e26819} (\bibinfo {year} {2022}{\natexlab{b}})}\BibitemShut
  {NoStop}%
\bibitem [{\citenamefont {Ronto}\ \emph {et~al.}(2023)\citenamefont {Ronto},
  \citenamefont {Jeszenszki}, \citenamefont {M\'atyus},\ and\ \citenamefont
  {Pollak}}]{RoJeMaPo23}%
  \BibitemOpen
  \bibfield  {author} {\bibinfo {author} {\bibfnamefont {M.}~\bibnamefont
  {Ronto}}, \bibinfo {author} {\bibfnamefont {P.}~\bibnamefont {Jeszenszki}},
  \bibinfo {author} {\bibfnamefont {E.}~\bibnamefont {M\'atyus}}, \ and\
  \bibinfo {author} {\bibfnamefont {E.}~\bibnamefont {Pollak}},\ }\href
  {\doibase 10.1103/PhysRevA.107.012204} {\bibfield  {journal} {\bibinfo
  {journal} {Phys. Rev. A}\ }\textbf {\bibinfo {volume} {107}},\ \bibinfo
  {pages} {012204} (\bibinfo {year} {2023})}\BibitemShut {NoStop}%
\bibitem [{\citenamefont {Ferenc}\ and\ \citenamefont
  {Mátyus}(2022)}]{FeMa22h3}%
  \BibitemOpen
  \bibfield  {author} {\bibinfo {author} {\bibfnamefont {D.}~\bibnamefont
  {Ferenc}}\ and\ \bibinfo {author} {\bibfnamefont {E.}~\bibnamefont
  {Mátyus}},\ }\href {\doibase https://doi.org/10.1016/j.cplett.2022.139734}
  {\bibfield  {journal} {\bibinfo  {journal} {Chem. Phys. Lett.}\ }\textbf
  {\bibinfo {volume} {801}},\ \bibinfo {pages} {139734} (\bibinfo {year}
  {2022})}\BibitemShut {NoStop}%
\bibitem [{\citenamefont {Li}\ \emph {et~al.}(2012)\citenamefont {Li},
  \citenamefont {Shao},\ and\ \citenamefont {Liu}}]{LiShLi12}%
  \BibitemOpen
  \bibfield  {author} {\bibinfo {author} {\bibfnamefont {Z.}~\bibnamefont
  {Li}}, \bibinfo {author} {\bibfnamefont {S.}~\bibnamefont {Shao}}, \ and\
  \bibinfo {author} {\bibfnamefont {W.}~\bibnamefont {Liu}},\ }\href {\doibase
  10.1063/1.3702631} {\bibfield  {journal} {\bibinfo  {journal} {J. Chem.
  Phys.}\ }\textbf {\bibinfo {volume} {136}},\ \bibinfo {pages} {144117}
  (\bibinfo {year} {2012})}\BibitemShut {NoStop}%
\bibitem [{\citenamefont {Almoukhalalati}\ \emph {et~al.}(2016)\citenamefont
  {Almoukhalalati}, \citenamefont {Knecht}, \citenamefont {Jensen},
  \citenamefont {Dyall},\ and\ \citenamefont {Saue}}]{AlKnJeDySa16}%
  \BibitemOpen
  \bibfield  {author} {\bibinfo {author} {\bibfnamefont {A.}~\bibnamefont
  {Almoukhalalati}}, \bibinfo {author} {\bibfnamefont {S.}~\bibnamefont
  {Knecht}}, \bibinfo {author} {\bibfnamefont {H.~J.~A.}\ \bibnamefont
  {Jensen}}, \bibinfo {author} {\bibfnamefont {K.~G.}\ \bibnamefont {Dyall}}, \
  and\ \bibinfo {author} {\bibfnamefont {T.}~\bibnamefont {Saue}},\ }\href
  {\doibase 10.1063/1.4959452} {\bibfield  {journal} {\bibinfo  {journal} {J.
  Chem. Phys.}\ }\textbf {\bibinfo {volume} {145}},\ \bibinfo {pages} {074104}
  (\bibinfo {year} {2016})}\BibitemShut {NoStop}%
\bibitem [{\citenamefont {Mittleman}(1981)}]{Mi81}%
  \BibitemOpen
  \bibfield  {author} {\bibinfo {author} {\bibfnamefont {M.~H.}\ \bibnamefont
  {Mittleman}},\ }\href {\doibase 10.1103/PhysRevA.24.1167} {\bibfield
  {journal} {\bibinfo  {journal} {Phys. Rev. A}\ }\textbf {\bibinfo {volume}
  {24}},\ \bibinfo {pages} {1167} (\bibinfo {year} {1981})}\BibitemShut
  {NoStop}%
\bibitem [{\citenamefont {Shao}\ \emph {et~al.}(2017)\citenamefont {Shao},
  \citenamefont {Li},\ and\ \citenamefont {Liu}}]{ShLiLi17}%
  \BibitemOpen
  \bibfield  {author} {\bibinfo {author} {\bibfnamefont {S.}~\bibnamefont
  {Shao}}, \bibinfo {author} {\bibfnamefont {Z.}~\bibnamefont {Li}}, \ and\
  \bibinfo {author} {\bibfnamefont {W.}~\bibnamefont {Liu}},\ }in\ \href
  {\doibase 10.1007/978-3-642-41611-8_7-1} {\emph {\bibinfo {booktitle}
  {Handbook of {{Relativistic Quantum Chemistry}}}}},\ \bibinfo {editor}
  {edited by\ \bibinfo {editor} {\bibfnamefont {W.}~\bibnamefont {Liu}}}\
  (\bibinfo  {publisher} {{Springer}},\ \bibinfo {address} {{Berlin,
  Heidelberg}},\ \bibinfo {year} {2017})\ pp.\ \bibinfo {pages}
  {481--496}\BibitemShut {NoStop}%
\bibitem [{\citenamefont {Tracy}\ and\ \citenamefont {Singh}(1972)}]{TrSi72}%
  \BibitemOpen
  \bibfield  {author} {\bibinfo {author} {\bibfnamefont {S.}~\bibnamefont
  {Tracy}}\ and\ \bibinfo {author} {\bibfnamefont {P.}~\bibnamefont {Singh}},\
  }\href@noop {} {\bibfield  {journal} {\bibinfo  {journal} {Stat. Neerl.}\
  }\textbf {\bibinfo {volume} {26}},\ \bibinfo {pages} {143} (\bibinfo {year}
  {1972})}\BibitemShut {NoStop}%
\bibitem [{\citenamefont {Dyall}\ and\ \citenamefont {{Faegri,
  Jr.}}(2007)}]{DyFaBook07}%
  \BibitemOpen
  \bibfield  {author} {\bibinfo {author} {\bibfnamefont {K.~G.}\ \bibnamefont
  {Dyall}}\ and\ \bibinfo {author} {\bibfnamefont {K.}~\bibnamefont {{Faegri,
  Jr.}}},\ }\href {\doibase 10.1093/oso/9780195140866.001.0001} {\emph
  {\bibinfo {title} {Introduction to Relativistic Quantum Chemistry}}}\
  (\bibinfo  {publisher} {Oxford University Press},\ \bibinfo {address} {New
  York},\ \bibinfo {year} {2007})\BibitemShut {NoStop}%
\bibitem [{\citenamefont {Kutzelnigg}(1984)}]{Ku84}%
  \BibitemOpen
  \bibfield  {author} {\bibinfo {author} {\bibfnamefont {W.}~\bibnamefont
  {Kutzelnigg}},\ }\href {\doibase 10.1002/qua.560250112} {\bibfield  {journal}
  {\bibinfo  {journal} {Int. J. Quant. Chem.}\ }\textbf {\bibinfo {volume}
  {25}},\ \bibinfo {pages} {107} (\bibinfo {year} {1984})}\BibitemShut
  {NoStop}%
\bibitem [{\citenamefont {Sun}\ \emph {et~al.}(2011)\citenamefont {Sun},
  \citenamefont {Liu},\ and\ \citenamefont {Kutzelnigg}}]{SuLiKu11}%
  \BibitemOpen
  \bibfield  {author} {\bibinfo {author} {\bibfnamefont {Q.}~\bibnamefont
  {Sun}}, \bibinfo {author} {\bibfnamefont {W.}~\bibnamefont {Liu}}, \ and\
  \bibinfo {author} {\bibfnamefont {W.}~\bibnamefont {Kutzelnigg}},\ }\href
  {\doibase 10.1007/s00214-010-0876-6} {\bibfield  {journal} {\bibinfo
  {journal} {Theor. Chim. Acta}\ }\textbf {\bibinfo {volume} {129}},\ \bibinfo
  {pages} {423} (\bibinfo {year} {2011})}\BibitemShut {NoStop}%
\bibitem [{\citenamefont {Szalewicz}\ and\ \citenamefont
  {Jeziorski}(2010)}]{SzJe10}%
  \BibitemOpen
  \bibfield  {author} {\bibinfo {author} {\bibfnamefont {K.}~\bibnamefont
  {Szalewicz}}\ and\ \bibinfo {author} {\bibfnamefont {B.}~\bibnamefont
  {Jeziorski}},\ }\href {\doibase 10.1080/00268976.2010.522206} {\bibfield
  {journal} {\bibinfo  {journal} {Mol. Phys.}\ }\textbf {\bibinfo {volume}
  {108}},\ \bibinfo {pages} {3091} (\bibinfo {year} {2010})}\BibitemShut
  {NoStop}%
\bibitem [{\citenamefont {Mátyus}(2019)}]{Ma19review}%
  \BibitemOpen
  \bibfield  {author} {\bibinfo {author} {\bibfnamefont {E.}~\bibnamefont
  {Mátyus}},\ }\href {\doibase 10.1080/00268976.2018.1530461} {\bibfield
  {journal} {\bibinfo  {journal} {Mol. Phys.}\ }\textbf {\bibinfo {volume}
  {117}},\ \bibinfo {pages} {590} (\bibinfo {year} {2019})}\BibitemShut
  {NoStop}%
\bibitem [{\citenamefont {Reiher}\ and\ \citenamefont
  {Wolf}(2015)}]{ReWoBook15}%
  \BibitemOpen
  \bibfield  {author} {\bibinfo {author} {\bibfnamefont {M.}~\bibnamefont
  {Reiher}}\ and\ \bibinfo {author} {\bibfnamefont {A.}~\bibnamefont {Wolf}},\
  }\href {https://onlinelibrary.wiley.com/doi/book/10.1002/9783527667550}
  {\emph {\bibinfo {title} {Relativistic Quantum Chemistry: The Fundamental
  Theory of Molecular Science, 2nd edition}}}\ (\bibinfo  {publisher}
  {Wiley-VCH},\ \bibinfo {address} {Weinheim},\ \bibinfo {year}
  {2015})\BibitemShut {NoStop}%
\bibitem [{\citenamefont {Kutzelnigg}(2012)}]{Ku12}%
  \BibitemOpen
  \bibfield  {author} {\bibinfo {author} {\bibfnamefont {W.}~\bibnamefont
  {Kutzelnigg}},\ }\href {\doibase 10.1016/j.chemphys.2011.06.001} {\bibfield
  {journal} {\bibinfo  {journal} {Chem. Phys.}\ }\textbf {\bibinfo {volume}
  {395}},\ \bibinfo {pages} {16} (\bibinfo {year} {2012})}\BibitemShut
  {NoStop}%
\bibitem [{\citenamefont {Liu}(2012)}]{Liu12}%
  \BibitemOpen
  \bibfield  {author} {\bibinfo {author} {\bibfnamefont {W.}~\bibnamefont
  {Liu}},\ }\href {\doibase 10.1039/C1CP21718F} {\bibfield  {journal} {\bibinfo
   {journal} {Phys. Chem. Chem. Phys.}\ }\textbf {\bibinfo {volume} {14}},\
  \bibinfo {pages} {35} (\bibinfo {year} {2012})}\BibitemShut {NoStop}%
\bibitem [{\citenamefont {Liu}\ \emph {et~al.}(2017)\citenamefont {Liu},
  \citenamefont {Shao},\ and\ \citenamefont {Li}}]{LiShLi17}%
  \BibitemOpen
  \bibfield  {author} {\bibinfo {author} {\bibfnamefont {W.}~\bibnamefont
  {Liu}}, \bibinfo {author} {\bibfnamefont {S.}~\bibnamefont {Shao}}, \ and\
  \bibinfo {author} {\bibfnamefont {Z.}~\bibnamefont {Li}},\ }in\ \href
  {\doibase 10.1007/978-3-642-41611-8_7-1} {\emph {\bibinfo {booktitle}
  {Handbook of {{Relativistic Quantum Chemistry}}}}},\ \bibinfo {editor}
  {edited by\ \bibinfo {editor} {\bibfnamefont {W.}~\bibnamefont {Liu}}}\
  (\bibinfo  {publisher} {{Springer}},\ \bibinfo {address} {{Berlin,
  Heidelberg}},\ \bibinfo {year} {2017})\ pp.\ \bibinfo {pages}
  {531--545}\BibitemShut {NoStop}%
\bibitem [{\citenamefont {Jeszenszki}\ and\ \citenamefont
  {Mátyus}(2024)}]{JeMa24CC}%
  \BibitemOpen
  \bibfield  {author} {\bibinfo {author} {\bibfnamefont {P.}~\bibnamefont
  {Jeszenszki}}\ and\ \bibinfo {author} {\bibfnamefont {E.}~\bibnamefont
  {Mátyus}},\ }\href@noop {} {\enquote {\bibinfo {title} {With-pair wave
  equation for the instantaneous coulomb--breit interaction: properties and
  application},}\ }\bibinfo {howpublished} {in preparation} (\bibinfo {year}
  {2024})\BibitemShut {NoStop}%
\bibitem [{\citenamefont {Wang}\ and\ \citenamefont {Yan}(2018)}]{WaYa18}%
  \BibitemOpen
  \bibfield  {author} {\bibinfo {author} {\bibfnamefont {L.~M.}\ \bibnamefont
  {Wang}}\ and\ \bibinfo {author} {\bibfnamefont {Z.-C.}\ \bibnamefont {Yan}},\
  }\href {\doibase 10.1103/PhysRevA.97.060501} {\bibfield  {journal} {\bibinfo
  {journal} {Phys. Rev. A}\ }\textbf {\bibinfo {volume} {97}},\ \bibinfo
  {pages} {060501(R)} (\bibinfo {year} {2018})}\BibitemShut {NoStop}%
\bibitem [{\citenamefont {Ferenc}\ and\ \citenamefont
  {M\'atyus}(2019)}]{FeMa19EF}%
  \BibitemOpen
  \bibfield  {author} {\bibinfo {author} {\bibfnamefont {D.}~\bibnamefont
  {Ferenc}}\ and\ \bibinfo {author} {\bibfnamefont {E.}~\bibnamefont
  {M\'atyus}},\ }\href {\doibase 10.1103/PhysRevA.100.020501} {\bibfield
  {journal} {\bibinfo  {journal} {Phys. Rev. A}\ }\textbf {\bibinfo {volume}
  {100}},\ \bibinfo {pages} {020501(R)} (\bibinfo {year} {2019})}\BibitemShut
  {NoStop}%
\bibitem [{\citenamefont {Ferenc}\ and\ \citenamefont
  {Mátyus}(2019)}]{FeMa19HH}%
  \BibitemOpen
  \bibfield  {author} {\bibinfo {author} {\bibfnamefont {D.}~\bibnamefont
  {Ferenc}}\ and\ \bibinfo {author} {\bibfnamefont {E.}~\bibnamefont
  {Mátyus}},\ }\href {\doibase 10.1063/1.5109964} {\bibfield  {journal}
  {\bibinfo  {journal} {J. Chem. Phys.}\ }\textbf {\bibinfo {volume} {151}},\
  \bibinfo {pages} {094101} (\bibinfo {year} {2019})}\BibitemShut {NoStop}%
\bibitem [{\citenamefont {Ferenc}\ \emph {et~al.}(2020)\citenamefont {Ferenc},
  \citenamefont {Korobov},\ and\ \citenamefont {M\'atyus}}]{FeKoMa20}%
  \BibitemOpen
  \bibfield  {author} {\bibinfo {author} {\bibfnamefont {D.}~\bibnamefont
  {Ferenc}}, \bibinfo {author} {\bibfnamefont {V.~I.}\ \bibnamefont {Korobov}},
  \ and\ \bibinfo {author} {\bibfnamefont {E.}~\bibnamefont {M\'atyus}},\
  }\href {\doibase 10.1103/PhysRevLett.125.213001} {\bibfield  {journal}
  {\bibinfo  {journal} {Phys. Rev. Lett.}\ }\textbf {\bibinfo {volume} {125}},\
  \bibinfo {pages} {213001} (\bibinfo {year} {2020})}\BibitemShut {NoStop}%
\bibitem [{\citenamefont {Saly}\ \emph {et~al.}(2023)\citenamefont {Saly},
  \citenamefont {Ferenc},\ and\ \citenamefont {Mátyus.}}]{SaFeMa22}%
  \BibitemOpen
  \bibfield  {author} {\bibinfo {author} {\bibfnamefont {E.}~\bibnamefont
  {Saly}}, \bibinfo {author} {\bibfnamefont {D.}~\bibnamefont {Ferenc}}, \ and\
  \bibinfo {author} {\bibfnamefont {E.}~\bibnamefont {Mátyus.}},\ }\href
  {\doibase 10.1080/00268976.2022.2163714} {\bibfield  {journal} {\bibinfo
  {journal} {Mol. Phys.}\ ,\ \bibinfo {pages} {e2163714}} (\bibinfo {year}
  {2023})}\BibitemShut {NoStop}%
\bibitem [{\citenamefont {Puchalski}\ \emph {et~al.}(2016)\citenamefont
  {Puchalski}, \citenamefont {Komasa}, \citenamefont {Czachorowski},\ and\
  \citenamefont {Pachucki}}]{PuKoCzPa16}%
  \BibitemOpen
  \bibfield  {author} {\bibinfo {author} {\bibfnamefont {M.}~\bibnamefont
  {Puchalski}}, \bibinfo {author} {\bibfnamefont {J.}~\bibnamefont {Komasa}},
  \bibinfo {author} {\bibfnamefont {P.}~\bibnamefont {Czachorowski}}, \ and\
  \bibinfo {author} {\bibfnamefont {K.}~\bibnamefont {Pachucki}},\ }\href
  {\doibase 10.1103/PhysRevLett.117.263002} {\bibfield  {journal} {\bibinfo
  {journal} {Phys. Rev. Lett.}\ }\textbf {\bibinfo {volume} {117}},\ \bibinfo
  {pages} {263002} (\bibinfo {year} {2016})}\BibitemShut {NoStop}%
\bibitem [{\citenamefont {{\v S}eba}(1988)}]{Se88}%
  \BibitemOpen
  \bibfield  {author} {\bibinfo {author} {\bibfnamefont {P.}~\bibnamefont {{\v
  S}eba}},\ }\href {\doibase 10.1007/BF00398170} {\bibfield  {journal}
  {\bibinfo  {journal} {Lett. Math. Phys.}\ }\textbf {\bibinfo {volume} {16}},\
  \bibinfo {pages} {51} (\bibinfo {year} {1988})}\BibitemShut {NoStop}%
\bibitem [{\citenamefont {Eyring}\ \emph {et~al.}(1944)\citenamefont {Eyring},
  \citenamefont {Walter},\ and\ \citenamefont {Kimball}}]{EyWaKi44}%
  \BibitemOpen
  \bibfield  {author} {\bibinfo {author} {\bibfnamefont {H.}~\bibnamefont
  {Eyring}}, \bibinfo {author} {\bibfnamefont {J.}~\bibnamefont {Walter}}, \
  and\ \bibinfo {author} {\bibfnamefont {G.}~\bibnamefont {Kimball}},\
  }\href@noop {} {\emph {\bibinfo {title} {{Quantum Chemistry, 1st ed.}}}}\
  (\bibinfo  {publisher} {John Wiley \& Sons},\ \bibinfo {address} {Canada},\
  \bibinfo {year} {1944})\BibitemShut {NoStop}%
\bibitem [{\citenamefont {L\"owdin}(1955)}]{Lo55}%
  \BibitemOpen
  \bibfield  {author} {\bibinfo {author} {\bibfnamefont {P.-O.}\ \bibnamefont
  {L\"owdin}},\ }\href {\doibase 10.1103/PhysRev.97.1474} {\bibfield  {journal}
  {\bibinfo  {journal} {Phys. Rev.}\ }\textbf {\bibinfo {volume} {97}},\
  \bibinfo {pages} {1474} (\bibinfo {year} {1955})}\BibitemShut {NoStop}%
\bibitem [{\citenamefont {Sucher}(1980)}]{Su80}%
  \BibitemOpen
  \bibfield  {author} {\bibinfo {author} {\bibfnamefont {J.}~\bibnamefont
  {Sucher}},\ }\href {\doibase 10.1103/PhysRevA.22.348} {\bibfield  {journal}
  {\bibinfo  {journal} {Phys. Rev. A}\ }\textbf {\bibinfo {volume} {22}},\
  \bibinfo {pages} {348} (\bibinfo {year} {1980})}\BibitemShut {NoStop}%
\bibitem [{\citenamefont {Sucher}(1983)}]{Su83}%
  \BibitemOpen
  \bibfield  {author} {\bibinfo {author} {\bibfnamefont {J.}~\bibnamefont
  {Sucher}},\ }\enquote {\bibinfo {title} {Foundations of the relativistic
  theory of many-electron systems},}\ in\ \href {\doibase
  10.1007/978-1-4613-3596-2_1} {\emph {\bibinfo {booktitle} {Relativistic
  Effects in Atoms, Molecules, and Solids}}},\ \bibinfo {editor} {edited by\
  \bibinfo {editor} {\bibfnamefont {G.~L.}\ \bibnamefont {Malli}}}\ (\bibinfo
  {publisher} {Springer US},\ \bibinfo {address} {Boston, MA},\ \bibinfo {year}
  {1983})\ pp.\ \bibinfo {pages} {1--53}\BibitemShut {NoStop}%
\bibitem [{\citenamefont {Sucher}(1984)}]{Su84}%
  \BibitemOpen
  \bibfield  {author} {\bibinfo {author} {\bibfnamefont {J.}~\bibnamefont
  {Sucher}},\ }\href {\doibase 10.1002/qua.560250103} {\bibfield  {journal}
  {\bibinfo  {journal} {Int. J. Quant. Chem.}\ }\textbf {\bibinfo {volume}
  {25}},\ \bibinfo {pages} {3} (\bibinfo {year} {1984})}\BibitemShut {NoStop}%
\end{thebibliography}
%


\clearpage

\setcounter{section}{0}
\renewcommand{\thesection}{S\arabic{section}}
\setcounter{subsection}{0}
\renewcommand{\thesubsection}{S\arabic{section}.\arabic{subsection}}

\setcounter{equation}{0}
\renewcommand{\theequation}{S\arabic{equation}}

\setcounter{table}{0}
\renewcommand{\thetable}{S\arabic{table}}

\setcounter{figure}{0}
\renewcommand{\thefigure}{S\arabic{figure}}

~\\[0.cm]
\begin{center}
\begin{minipage}{0.8\linewidth}
\centering
\textbf{Supplementary Information to } \\[0.25cm]

\textbf{%
One-Particle Operator Representation over Two-Particle Basis Sets for Relativistic QED Computations
}
\end{minipage}
~\\[0.5cm]
\begin{minipage}{0.6\linewidth}
\centering

Péter Hollósy,$^1$ P\'eter Jeszenszki,$^1$ and Edit M\'atyus$^{1,\ast}$ \\[0.15cm]

$^1$~\emph{ELTE, Eötvös Loránd University, Institute of Chemistry, 
Pázmány Péter sétány 1/A, Budapest, H-1117, Hungary} \\[0.15cm]
$^\ast$ edit.matyus@ttk.elte.hu \\
\end{minipage}
~\\[0.15cm]
(Dated: \today)
\end{center}

\clearpage

\begin{table}[h!]
  \caption{%
    Data for Figure~4 of the manuscript: Projected Dirac--Coulomb energy of the helium atom ground state as a function of the $E_\mathrm{th}$ threshold energy used to define the cutting projector. All energies are in $E_\text{h}$ units.
  }  
  \centering
  \begin{tabular}{@{}r c c r r@{}}
    \hline\hline
      $E_\text{th}$ &   & 
      $E_\cutting$ & 
      $(E_\cutting-E_{\text{\honehtwo}}) 10^9$ & 
      $N[\%]$ \\
    \hline
        --4.1   &    & --2.903 856 631 6 & --0.002 & 26.0 \\
        --6259  &    & --2.903 856 631 6 & --0.002 & 26.0 \\
        --12514 &    & --2.903 856 631 6 & --0.001 & 26.1 \\
        --18769 &    & --2.903 856 631 6 & --0.002 & 26.3 \\
        --25025 &    & --2.903 856 631 6 & --0.002 & 26.6 \\
        --31282 &    & --2.903 856 630 5 &   1.045 & 27.4 \\
        --37528 &    & --2.903 856 568 5 &  63.041 & 39.2 \\
        --43794 &    & --2.903 856 508 2 & 123.306 & 72.6 \\
        --56307 &    & --2.903 856 508 2 & 123.158 & 73.7 \\
        --62563 &    & --2.903 856 508 3 & 123.256 & 73.9 \\
        --68820 &    & --2.903 856 508 4 & 123.199 & 74.0 \\
        --75076 &    & --2.903 856 508 5 & 123.097 & 74.0 \\
        --75200 &    & --2.903 856 508 3 & 123.238 & 85.9 \\
        --100000 &   & --2.903 856 507 5 & 124.054 & 98.0 \\
        --200000 &   & --2.903 856 505 3 & 126.245 & 99.3 \\
        --500000 &   & --2.903 856 504 6 & 126.946 & 99.8 \\
        --1000000 &  & --2.903 856 504 0 & 127.533 & 99.9 \\
        \hline
        no projection &    & --2.903 856 504 7 & 126.916 & 100 \\
    \hline\hline
    \end{tabular}
    \label{tab_Threshold}
\end{table}

\begin{table}[h!]
  \caption{%
  He atom no-pair Dirac--Coulomb energy (ground state, singlet basis sector). 
  Comparison of the cutting, the \honehtwo, and the punching projectors. The punching(CCR) and punching(\honehtwo) projector results are identical in this example.
  The energies and the $\Delta E_\text{\honehtwo}$ = $E_\cutting-E_{\text{\honehtwo}}$ and $\Delta E_\punching$ = $E_\cutting-E_{\punching}$ energy differences are in $\Eh$ units.  
  $N$ is the number of ECG functions.
  Quadruple precision arithmetic is used in all computations.
    }
    \centering
     \begin{tabular}{@{}ccc crr@{}}
    \hline\hline
    $N$ &   & 
    $E_\cutting$ & 
    $E_{\text{\honehtwo}}$ & 
    \multicolumn{1}{c}{$\Delta E_\text{\honehtwo}$} & 
    \multicolumn{1}{c}{$\Delta E_\text{punching}$} \\
    \hline
        10   &   & --2.898 070 389 8 & --2.898 070 389 8 & --$4\cdot10^{-11}$ & \\
        100  &   & --2.903 856 456 9 & --2.903 856 458 5 &   $2\cdot10^{-9}$  & --$7\cdot10^{-18}$ \\
        200  &   & --2.903 856 618 3 & --2.903 856 618 2 & --$8\cdot10^{-11}$ & --$7\cdot10^{-15}$ \\
        300  &   & --2.903 856 629 7 & --2.903 856 629 7 &   $5\cdot10^{-13}$ & --$5\cdot10^{-11}$ \\
        400  &   & --2.903 856 631 3 & --2.903 856 631 3 & --$6\cdot10^{-12}$ & --$1\cdot10^{-14}$ \\
        500  &   & --2.903 856 631 5 & --2.903 856 631 6 &   $3\cdot10^{-12}$ & --$1\cdot10^{-8}$ \\
        700  &   & --2.903 856 631 6 & --2.903 856 631 6 &   $5\cdot10^{-12}$ & --$3\cdot10^{-11}$ \\
    %
    \hline\hline
    \end{tabular}
    \label{tab:HeDC}
\end{table}

\begin{table}[h!]
    \caption{%
  He atom no-pair Dirac--Coulomb--Breit energy (ground state, singlet basis sector).    
  Comparison of the cutting, the \honehtwo, and the punching projectors. 
  The punching(CCR) and punching(\honehtwo) projector results are identical in this example.
  The energies and the $\Delta E_\text{\honehtwo}$ = $E_\cutting-E_{\text{\honehtwo}}$ and $\Delta E_\punching$ = $E_\cutting-E_{\punching}$ energy differences are in $\Eh$ units.  
  $N$ is the number of ECG functions.
  Quadruple precision arithmetic is used in all computations.
}
    \centering
     \begin{tabular}{@{}c c c c l l@{}}
    \hline\hline
        $N$  &   & 
        $E_\cutting$        & 
        $E_{{\text{\honehtwo}}}$  & 
        $\Delta E_\text{\honehtwo}$ & 
        $\Delta E_\punching$ \\
    \hline
        10   &   & --2.898 031 259 9 & --2.898 031 259 8 & --$6\cdot10^{-11}$ &  \\
        100  &   & --2.903 827 907 1 & --2.903 827 903 4 & --$4\cdot10^{-9}$  & --$5\cdot10^{-17}$ \\
        200  &   & --2.903 828 103 0 & --2.903 828 102 6 & --$4\cdot10^{-10}$ & --$1\cdot10^{-14}$ \\
        300  &   & --2.903 828 118 3 & --2.903 828 118 1 & --$2\cdot10^{-10}$ & --$7\cdot10^{-11}$ \\
        400  &   & --2.903 828 120 7 & --2.903 828 120 6 & --$1\cdot10^{-10}$ & --$2\cdot10^{-14}$ \\
        500  &   & --2.903 828 121 1 & --2.903 828 121 0 & --$1\cdot10^{-10}$ & --$2\cdot10^{-8}$\\
        700  &   & --2.903 828 121 1 & --2.903 828 121 0 & --$1\cdot10^{-10}$ & --$3\cdot10^{-11}$\\
    %
    \hline\hline
    \end{tabular}
    \label{tab:HeDCB}
\end{table}

\begin{table}[h!]
    \caption{%
  Ar$^{16+}$ atom no-pair Dirac--Coulomb energy (ground state, singlet basis sector).    
  Comparison of the cutting, the \honehtwo, the punching(\honehtwo), and the punching(CCR) projectors. 
  The energies and the $\Delta E_\text{\honehtwo}$ = $E_\cutting-E_{\text{\honehtwo}}$ and $\Delta E_\punching$ = $E_\cutting-E_{\punching}$ energy differences are in $\Eh$ units.  
  $N$ is the number of ECG functions.
  Quadruple precision arithmetic is used in all computations.  
    }
    \centering
     \begin{tabular}{@{}c c c c l l l@{}}
    \hline\hline
    $N$  &   & $E_\cutting$        & 
    $E_\text{\honehtwo}$  & 
    $\Delta E_\text{\honehtwo}$ & 
    $\Delta E_\text{punching}^\text{CCR}$ & 
    $\Delta E_\text{punching}^\text{\honehtwo}$ \\
    \hline
        100  &   & --314.246 075 105 6 & --314.246 074 912 8 & --$2\cdot10^{-7}$ & --$3\cdot10^{-5}$ & --$3\cdot10^{-5}$\\
        200  &   & --314.246 094 310 9 & --314.246 094 296 5 & --$1\cdot10^{-8}$ & --$1\cdot10^{-4}$ & --$9\cdot10^{-5}$ \\
        300  &   & --314.246 098 798 8 & --314.246 098 658 5 & --$1\cdot10^{-7}$ & --$4\cdot10^{-6}$ & --$3\cdot10^{-6}$\\
        400  &   & --314.246 103 139 2 & --314.246 103 099 5 & --$4\cdot10^{-8}$ & --$3\cdot10^{-5}$ & --$3\cdot10^{-5}$\\
        500  &   & --314.246 103 541 7 & --314.246 103 498 5 & --$4\cdot10^{-8}$ & --$8\cdot10^{-6}$ & --$2\cdot10^{-5}$\\
        600  &   & --314.246 103 563 3 & --314.246 103 520 6 & --$4\cdot10^{-8}$ & --$1\cdot10^{-5}$ & --$4\cdot10^{-5}$ \\
        700  &   & --314.246 103 987 9 & --314.246 103 954 2 & --$3\cdot10^{-8}$ & --$1\cdot10^{-5}$ & --$1\cdot10^{-5}$\\
        800  &   & --314.246 104 156 6 & --314.246 104 129 0 & --$3\cdot10^{-8}$ & --$8\cdot10^{-6}$ & --$8\cdot10^{-6}$ \\
    %
    \hline\hline
    \end{tabular}
    \label{tab:ArDC}
\end{table}

\begin{table}[h!]
  \caption{%
  Ar$^{16+}$ atom no-pair Dirac--Coulomb--Breit energy (ground state, singlet basis sector).    
  Comparison of the cutting, the \honehtwo, the punching(\honehtwo), and the punching(CCR) projectors. 
  The energies and the $\Delta E_\text{\honehtwo}$ = $E_\cutting-E_{\text{\honehtwo}}$ and $\Delta E_\punching$ = $E_\cutting-E_{\punching}$ energy differences are in $\Eh$ units.  
  $N$ is the number of ECG functions.
  Quadruple precision arithmetic is used in all computations.  
  }
    \centering
     \begin{tabular}{@{}c c c c l l l@{}}
    \hline\hline
    $N$  &   & 
    $E_\cutting$ & 
    $E_\text{\honehtwo}$ & 
    $\Delta E_\text{\honehtwo}$ & 
    $\Delta E_\punching^\text{CCR}$ & 
    $\Delta E_\punching^\text{\honehtwo}$ \\
    \hline
        100  &   & --314.176 869 342 3 & --314.176 867 488 3 & --$2\cdot10^{-6}$ & --$4\cdot10^{-5}$ & --$4\cdot10^{-5}$ \\
        200  &   & --314.176 909 642 0 & --314.176 909 395 8 & --$2\cdot10^{-7}$ & --$1\cdot10^{-4}$ & --$1\cdot10^{-4}$ \\
        300  &   & --314.176 914 174 9 & --314.176 913 941 5 & --$2\cdot10^{-7}$ & --$1\cdot10^{-5}$ & --$7\cdot10^{-6}$ \\
        400  &   & --314.176 918 552 2 & --314.176 918 405 7 & --$2\cdot10^{-7}$ & --$5\cdot10^{-5}$ & --$5\cdot10^{-5}$ \\
        500  &   & --314.176 918 980 8 & --314.176 918 865 9 & --$1\cdot10^{-7}$ & --$3\cdot10^{-5}$ & --$2\cdot10^{-5}$  \\
        600  &   & --314.176 919 016 0 & --314.176 918 889 6 & --$1\cdot10^{-7}$ & --$4\cdot10^{-5}$ & --$1\cdot10^{-4}$ \\
        700  &   & --314.176 919 461 6 & --314.176 919 327 5 & --$1\cdot10^{-7}$ & --$2\cdot10^{-5}$ & --$2\cdot10^{-5}$ \\
        800  &   & --314.176 919 676 5 & --314.176 919 546 5 & --$1\cdot10^{-7}$ & --$2\cdot10^{-5}$ & --$1\cdot10^{-5}$ \\
    %
    \hline\hline
    \end{tabular}
    \label{tab:ArDCB}
\end{table}

\end{document}